# Inconsistency Robustness in Foundations: Mathematics self proves its own Consistency and Other Matters

**Carl Hewitt**

***This article is dedicated to Alonzo Church, Stanisław Jaśkowski, Ludwig Wittgenstein, and Ernst Zermelo.***

## Abstract

Inconsistency Robustness is performance of information systems with pervasively inconsistent information. Inconsistency Robustness of the community of professional mathematicians is their performance repeatedly repairing contradictions over the centuries. In the Inconsistency Robustness paradigm, deriving contradictions have been a progressive development and not "game stoppers." Contradictions can be helpful instead of being something to be "swept under the rug" by denying their existence, which has been repeatedly attempted by Establishment Philosophers (beginning with some Pythagoreans). Such denial has delayed mathematical development. This article reports how considerations of Inconsistency Robustness have recently influenced the foundations of mathematics for Computer Science continuing a tradition developing the sociological basis for foundations.[1]

Classical Direct Logic is a foundation of mathematics for Computer Science, which has a foundational theory (for convenience called "Mathematics") that can be used in any other theory. A bare turnstile is used for Mathematics so that $\vdash\Psi$ means that $\Psi$ is a mathematical proposition that is a theorem of Mathematics and $\Phi\vdash\Psi$ means that $\Psi$ can be inferred from $\Phi$ in Mathematics.

The current common understanding is that Gödel proved "Mathematics cannot prove its own consistency, if it is consistent." However, the consistency of mathematics can be proved by a simple argument using standard rules of Mathematics including the following:

- rule of Proof by Contradiction, *i.e.*, $(\neg\Phi\Rightarrow(\Theta\wedge\neg\Theta))\vdash\Phi$
- and the rule of Soundness, *i.e.*, $(\vdash\Phi)\Rightarrow\Phi$

*Formal Proof.* By definition,

Consistent$\Leftrightarrow\neg\exists[\Psi\text{:Proposition}]\rightarrow\vdash(\Psi\wedge\neg\Psi)$. By Existential Elimination, there is some proposition $\Psi_0$ such that $\neg$Consistent $\Rightarrow\vdash(\Psi_0\wedge\neg\Psi_0)$ which by Soundness and transitivity of implication means $\neg$Consistent$\Rightarrow(\Psi_0\wedge\neg\Psi_0)$. Substituting for $\Phi$ and $\Theta$, in the rule for Proof by Contradiction, it follows that $(\neg\text{Consistent}\Rightarrow(\Psi_0\wedge\neg\Psi_0))\vdash$ Consistent. Thus, $\vdash$ Consistent.

The above theorem means that consistency is deeply embedded in the architecture of classical mathematics. Please note the following points: **The above argument formally mathematically proves the theorem that mathematics is consistent** and that **it is *not* a premise of the theorem that mathematics is consistent.** Classical mathematics was designed for consistent axioms and consequently the rules of classical mathematics can be used to prove consistency regardless of the axioms, *e.g.*, Euclidean geometry.

The above proof means that “Mathematics is consistent” is a theorem in Classical Direct Logic. This means that the usefulness of Classical Direct Logic depends crucially on the consistency of Mathematics. Good evidence for the consistency of Mathematics comes from the way that Classical Direct Logic avoids the known paradoxes. Humans have spent millennia devising paradoxes.

Having a powerful system like Direct Logic is important in computer science because computers must be able to formalize all logical inferences (including inferences about their own inference processes) without requiring recourse to human intervention. Any inconsistency in Classical Direct Logic would be a potential security hole because it could be used to cause computer systems to adopt invalid conclusions.

**The recently developed self-proof of consistency (above) shows that the current common understanding that Gödel proved “Mathematics cannot prove its own consistency, if it is consistent” is inaccurate**.[i]

Wittgenstein long ago showed that contradiction in mathematics results from the kind of “self-referential”[ii] sentence that Gödel used in his argument that Mathematics cannot prove its own consistency. However, using a typed grammar for mathematical sentences, it can be proved that the kind “self-referential” sentence that Gödel used in his argument cannot be constructed because the required fixed point that Gödel used to construct the “self-referential” sentence does not exist. In this way, consistency of mathematics is preserved without giving up power.[2]

---

[i] Four years after Gödel published his results for Principia Mathematica, [Church 1935, Turing 1936] published the first valid proof that the mathematical theory *Principia* is inferentially undecidable (*i.e.* there is a proposition $\Psi$ such that $\nvdash_{\mathit{Principia}}\Psi$ and $\nvdash_{\mathit{Principia}}\neg\Psi$) because provability in *Principia* is computationally undecidable (provided that the theory *Principia* is consistent).

[ii] There seem to be no practical uses for “self-referential” propositions in the mathematical foundations of Computer Science.

## Mathematical Foundation for Computer Science

Computer Science brought different concerns and a new perspective to mathematical foundations including the following requirements:[3] [Arabic numeral superscripts refer to endnotes at the end of this article]

- provide powerful inference machinery so that arguments (proofs) can be short and understandable and all logical inferences can be formalized
- establish standard foundations so people can join forces and develop common techniques and technology
- incorporate axioms thought to be consistent by the overwhelming consensus of working professional mathematicians, e.g., natural numbers [Dedekind 1888, Peano 1889], real numbers [Dedekind 1888], sets of sets of integers, real, strings, *etc*.
- facilitate inferences about the mathematical foundations used by computer systems.

Classical Direct Logic is a foundation of mathematics for Computer Science, which has a foundational theory (for convenience called "Mathematics") that can be used in any other theory. A bare turnstile is used for Mathematics so that $\vdash\Psi$ means that $\Psi$ is a mathematical proposition that is a theorem of Mathematics and $\Phi\vdash\Psi$ means that $\Psi$ can be inferred from $\Phi$ in Mathematics.

## Mathematics self proves its own consistency

A mathematically significant idea involves:

> *"…a very high degree of unexpectedness, combined with inevitability and economy."* [Hardy 1940]

The following rules are fundamental to classical mathematics:

- Proof by Contradiction, *i.e.* $(\neg\Phi\Rightarrow(\Theta\wedge\neg\Theta))\vdash\Phi$, which says that a proposition can be proved by showing that it implies a contradiction.
- Soundness, *i.e.* $(\vdash\Phi)\Rightarrow\Phi$, which says that a theorem can be used in a proof.[4]

**Theorem:** Mathematics self proves its own consistency.[5]

*Formal Proof*[6] By definition, $\neg\text{Consistent}\Leftrightarrow\exists[\Psi\text{:Proposition}]\rightarrow\vdash(\Psi\wedge\neg\Psi)$.[7] By the rule of Existential Elimination, there is some proposition $\Psi_0$ such that $\neg\text{Consistent}\Rightarrow\vdash(\Psi_0\wedge\neg\Psi_0)$ which by the rule of Soundness and transitivity of implication means $\neg\text{Consistent}\Rightarrow(\Psi_0\wedge\neg\Psi_0)$. Substituting for Φ and Θ, in the rule for Proof by Contradiction, we have $(\neg\text{Consistent}\Rightarrow(\Psi_0\wedge\neg\Psi_0)) \vdash \text{Consistent}$. Thus, $\vdash \text{Consistent}$.

A Natural Deduction[i] proof is given below:

**1)** $\neg$Consistent // hypothesis to derive a contradiction **just in this subargument**

**2)** $\exists[\Psi\text{:Proposition}]\rightarrow \vdash(\Psi\wedge\neg\Psi)$ // definition of inconsistency using **1)**

**3)** $\vdash(\Psi_0\wedge\neg\Psi_0)$ // rule of Existential Elimination using **2)**

**4)** $\Psi_0\wedge\neg\Psi_0$ // rule of Soundness using **3)**

$\vdash$ Consistent // rule of Proof by Contradiction using **1)** and **4)**

**Natural Deduction Proof of Consistency of Mathematics**

[i] [Jaśkowski 1934] developed Natural Deduction *cf.* [Barker-Plummer, Barwise, and Etchemendy 2011]

Please note the following points:

- The above argument formally mathematically proves that mathematics is consistent and that **it is not a premise of the theorem that mathematics is consistent.**[8]
- Classical mathematics was designed for consistent axioms and consequently the rules of classical mathematics can be used to prove consistency regardless of other axioms.[9]

The above proof means that "Mathematics is consistent" is a theorem in Classical Direct Logic. This means that the usefulness of Classical Direct Logic depends crucially on the consistency of Mathematics.[10] Good evidence for the consistency of Mathematics comes from the way that Classical Direct Logic avoids the known paradoxes. Humans have spent millennia devising paradoxes.

Computer Science needs very strong foundations for mathematics so that computer systems are not handicapped. It is important not to have inconsistencies in mathematical foundations of Computer Science because they represent security vulnerabilities.

The recently developed self-proof of consistency (above) shows that the current common understanding that Gödel proved "Mathematics cannot prove its own consistency, if it is consistent" is inaccurate. But the situation is even more interesting because Wittgenstein more than a half-century ago showed that contradiction in mathematics results from the kind of "self-referential" sentence that Gödel used in his proof. Fortunately, using a typed grammar for mathematical sentences, it can be proved that the kind "self-referential" sentence that Gödel used in his proof cannot be constructed because required fixed points do not exist. Consequently, using a typed grammar, consistency of mathematics can be preserved without giving up power.

Formal typed grammars had not yet been invented when Gödel and other philosophers developed the First-order Thesis that weakened the foundations of mathematics so that, as expressed, "self-referential" propositions do not infer contradiction.[11] The weakened foundations (based on first-order logic) enabled some limited meta-mathematical theorems to be proved. However, as explained in this article, the weakened foundations are cumbersome, unnatural, and unsuitable as the mathematical foundation for Computer Science.

## Monster-Barring

> "*But why accept the counterexample? ... Why should the theorem give way...? It is the 'criticism' that should retreat.... It is a monster, a pathological case, not a counterexample*."
> Delta, student in [Lakatos, 1976, pg. 14].

The Euler formula for polyhedra is Vertices-Edges+Faces=2, which can be proved in a variety of different ways.

But the hollow cube below is a counterexample because Vertices-Edges+Faces=4.

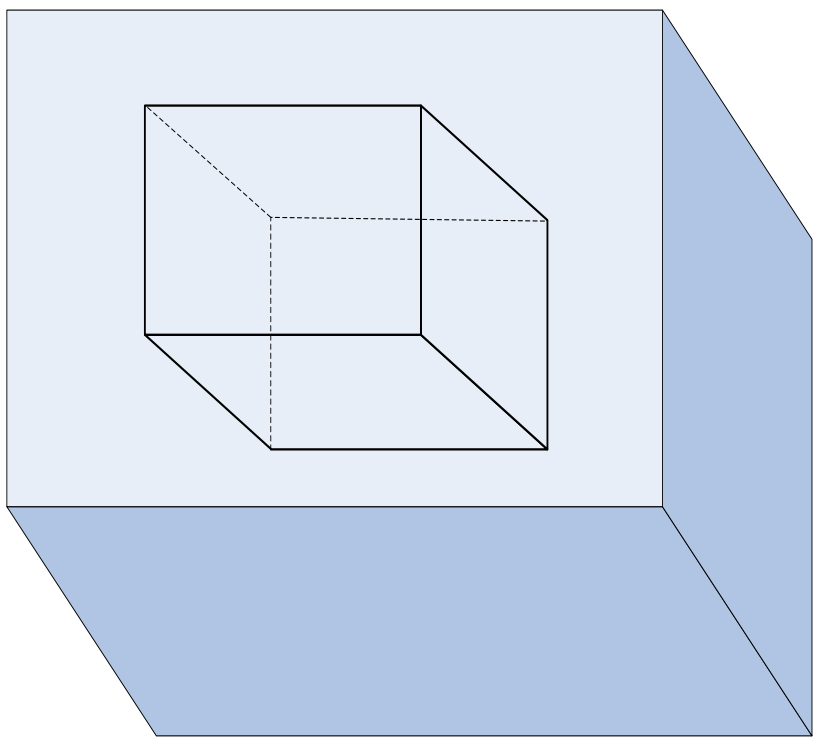

**Counterexample to Euler's Formula**

In the face of this counterexample, it becomes important to characterize polyhedra more rigorously. For example,

- A Regular solid
- A convex solid with polyhedral faces
- A surface consisting of a system of polygons
- *etc.*

Lakatos has called this strategy "*monster-barring*."

## Wittgenstein: "self-referential" propositions lead to inconsistency in mathematics

> *All truth passes through three stages:*
> *First, it is ridiculed.*
> *Second, it is violently opposed.*
> *Third, it is accepted as being self-evident.*
> Arthur Schopenhauer (1788-1860)

Early on, Wittgenstein correctly noted that Gödel's "self-referential" proposition infers inconsistency in mathematics:[i]

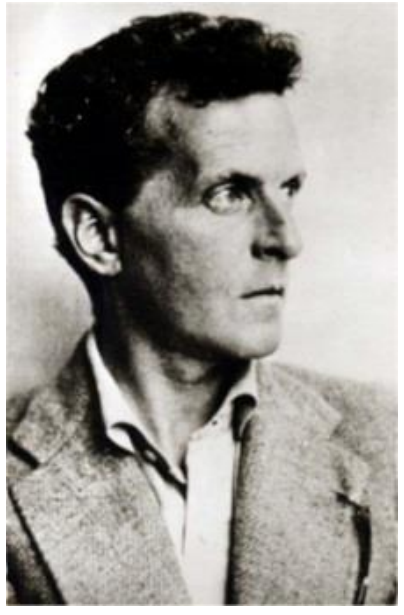

Ludwig Wittgenstein

> *Let us suppose I prove*[ii] *the improvability (in Russell's system) of* [Gödel's "self-referential" proposition[iii]] *P*; [$\vdash\nvdash P$ where $P \Leftrightarrow \nvdash P$] *then by this proof I have proved P* [$\vdash P$].
>
> *Now if this proof were one in Russell's system* [$\vdash\vdash P$]*—I should in this case have proved at once that it belonged* [$\vdash P$] *and did not belong* [$\vdash\neg P$ because $\neg P \Leftrightarrow \vdash P$] *to Russell's system.*
>
> *But there is a contradiction here!*

According to [Monk 2007]:[12]
*Wittgenstein hoped that his work on mathematics would have a cultural impact, that it would threaten the attitudes that prevail in logic, mathematics and the philosophies of them. On this measure it has been a spectacular failure.*

Unfortunately, recognition of the worth of Wittgenstein's work on mathematics came long after his death. For decades, professional work logicians mistakenly believed that they had been completely victorious over Wittgenstein.

## *contra* Gödel *et. al*

> *"Men… think in herds …*
> *they only recover their senses slowly, and one by one."*
> Charles Mackay

---

[i] Wittgenstein in 1937 published in Wittgenstein 1956, p. 50e and p. 51e]

[ii] Wittgenstein was granting the supposition that Gödel had proved inferential undecidability (sometimes called "incompleteness") of Russell's system, *e.g.*, $\vdash\nvdash P$. However, inferential undecidability is easy to prove using the "self-referential" proposition *P*:

*Proof.* Suppose to obtain a contradiction that $\vdash P$. Both of the following can be inferred:

1) $\vdash \nvdash P$ from the hypothesis because $P \Leftrightarrow \nvdash P$
2) $\vdash \vdash P$ from the hypothesis by Adequacy.

But 1) and 2) are a contradiction. Consequently, $\vdash\nvdash P$ follows from proof by contradiction.

[iii] constructed using a fixed point exploiting an untyped grammar of mathematics

That mathematics self proves its own consistency contradicts the result [Gödel 1931] using a "self-referential" proposition[i] that mathematics cannot prove its own consistency.

One resolution is not to have "self-referential" propositions.[ii] This can be achieved by carefully arranging the rules using a properly constructed grammar so that "self-referential" propositions cannot be constructed as shown below.[iii] The basic idea is to use types [Russell 1908, Church 1940] to construct propositions from other propositions so that fixed points do not exist and consequently cannot be used to construct "self-referential" propositions.

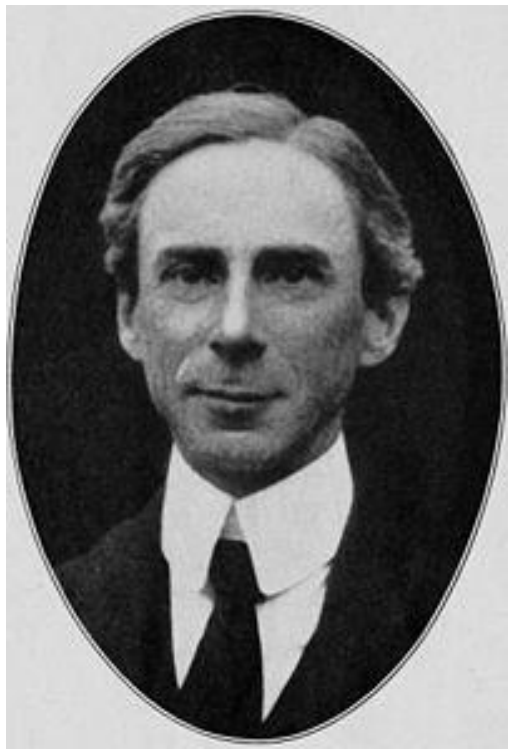
Bertrand Russell

However, there is a crucial difference between how Russell used types and the method used in Direct Logic. Russell attempted to use types as the fundamental mechanism for preventing inconsistencies by restricting the domain of mathematics to object that can be described by a strict hierarchical type system. However, he ran into trouble because his type mechanism was too strict and prevented ordinary mathematical reasoning.[iv]

In this paper, types are used to prevent the construction of "self-referential" sentences and to provide the foundations for sets. The difficulties encountered by Russell are avoided as follows:

- having integers[13] as primitive
- constructing sets from the characteristic functions of typed functions
- types are used to resolve the usual paradoxes with sets, *e.g.*, there is no set of all sets, *etc*.[14]

[i] constructed using a fixed point operator exploiting an untyped grammar for sentences

[ii] There do not seem to be any practical uses of "self-referential" propositions in the mathematical foundations of Computer Science.

[iii] It is important to note that disallowing "self-referential" propositions does not place restrictions on recursion in computation, *e.g.*, the Actor Model, untyped lambda calculus, *etc.*

[iv] In order to be able to carry out ordinary mathematical reasoning, Russell introduced an (unmotivated) patch called "ramified types" that collapsed the type hierarchy.

The above approach provides a very usable foundation for ordinary mathematical reasoning. Combining types and sets as the foundation has the advantage of using the strengths of each without the limitations of trying to use just one because each can be used to make up for the limitations of the other. The key idea is compositionality, *i.e.*, composing new entities from others. Types can be composed from other types and sets can be composed from other sets.[i]

## Classical Direct Logic

> *I suspect there are few today who share ...* [the] *belief that there should be a single overarching theory embracing all of mathematics.* [Dowson 2006]

Classical Direct Logic must meet the following challenges:

- *Consistent* to avoid security holes
- *Powerful* so that computer systems can formalize all logical inferences
- *Principled* so that it can be easily learned by software engineers
- *Coherent* so that it hangs together without a lot of edge cases
- *Intuitive* so that humans can follow computer system reasoning
- *Comprehensive* to accommodate all forms of logical argumentation
- *Inconsistency Robust* to be applicable to pervasively inconsistent theories of practice with
    - Inconsistency Robust Direct Logic for logical inference about inconsistent information
    - Classical Direct Logic for Mathematics used in inconsistency-robust theories

In Direct Logic, unrestricted recursion is allowed in programs by using recursive definitions.

---

[i] Compositionality avoids standard foundational paradoxes. For example, Direct Logic composes sentences from others using types so there are no “self-referential” propositions.

There are uncountably many Actors.[15] For example, Real.[ ] can output any real number[i] between 0 and 1 where[ii]

Real.[ ] ≡ [(0 **either** 1), ⩛**Postpone** Real.[ ]]

where

- Real.[ ] is the result of sending the Actor Real the message [ ]
- (0 **either** 1) is the nondeterministic choice of 0 or 1,
- [ *first*, ⩛*rest*] is the list that begins with *first* and whose remainder is *rest*, and
- **Postpone** *expression* delays execution of *expression* until the value is needed.

There are uncountably many propositions (because there is a different proposition for every real number). Consequently, there are propositions that are not the abstraction of any element of a denumerable set of sentences. For example, p ≡ [x:ℝ]⟶ ([y:ℝ]⟶ (y=x)) defines a different predicate p[x] for each real number x, which holds for only one real number, namely x.[iii]

It is important to distinguish between strings, sentences, and propositions. Some strings can be parsed into sentences (*i.e.* grammar tree structures), which can be abstracted into propositions that can be asserted. Furthermore, grammar terms (*i.e.* tree structures) can be abstracted into Actors (*i.e.* objects in mathematics).

Direct Logic distinguishes between concrete **sentences** and abstract **propositions**.[16] For example, the follow sentence is a Latin parse of the string "Gallia est omnis divisa in partes tres.":

$\ulcorner$"Gallia est omnis divisa in partes tres."$\urcorner_{Latin}$

On the other hand, the proposition that *All of Gaul is divided into three parts* was believed by Caesar.[17]

A sentence s can be **abstracted** ($\lfloor s \rfloor_{\mathcal{T}}$)[18] as a proposition in a theory $\mathcal{T}$. For example,

$\lfloor\ulcorner$"Gallia est omnis divisa in partes tres."$\urcorner_{Latin}\rfloor_{English}$

⇾ *All of Gaul is divided into three parts*

---

[i] using binary representation.

[ii] Typically, a result returned by the non-deterministic procedure Real is not computable in the sense there is no computable deterministic procedure that can compute its digits.

[iii] For example (p[3])[y] holds if and only if y=3.

Also,

⌊⌈"Gallia est omnis divisa in partes tres."⌉$_{Latin}$⌋$_{Spanish}$

→ *Toda Galia está dividida en tres partes*[i]

Abstraction and parsing are becoming increasingly important in software engineering. *e.g.,*

- The execution of code can be dynamically checked against its documentation. Also Web Services can be dynamically searched for and invoked on the basis of their documentation.
- Use cases can be inferred by specialization of documentation and from code by automatic test generators and by model checking.
- Code can be generated by inference from documentation and by generalization from use cases.

**Abstraction and parsing are needed for large software systems so that that documentation, use cases, and code can mutually speak about what has been said and their relationships.**

In mathematics:

**Proposition**

*e.g.* ∀[n:ℕ]→ ∃[m:ℕ]→ m>n

**Sentence**

*e.g.*⌈"∀[n:ℕ]→ ∃[m:ℕ]→ m>n"⌉

**String**

*e.g.* "∀[n:ℕ]→ ∃[m:ℕ]→ m>n"

---

[i] Spanish for all of Gaul is divided in three parts.

In Direct Logic, a sentence is a grammar tree (analogous to the ones used by linguists). Such a grammar tree has terminals that can be constants. And there are uncountably many constants, *e.g.*, the real numbers:

> The sentence ⌈3.14159... < 3.14159... + 1⌉ is impossible to obtain by parsing a string (where 3.14159... is an Actor[i] for the transcendental real number that is the ratio of a circle's circumference to its diameter). The issue is that there is no string which when parsed is
>
> ⌈3.14159... < 3.14159... + 1⌉
>
> Of course, because the digits of 3.14159... are computable, there is an $\mathsf{term}_1$ such that $\lfloor \mathsf{term}_1 \rfloor$[ii] = 3.14159... that can be used to create the sentence $\lceil \mathsf{term}_1 < \mathsf{term}_1 + 1 \rceil$.
>
> However the sentence $\lceil \mathsf{term}_1 < \mathsf{term}_1 + 1 \rceil$ is not the same as ⌈3.14159... < 3.14159... + 1⌉ because it does not have the same vocabulary and it is a much larger sentence that has many terminals whereas ⌈3.14159... < 3.14159... + 1⌉ has just 3 terminals:

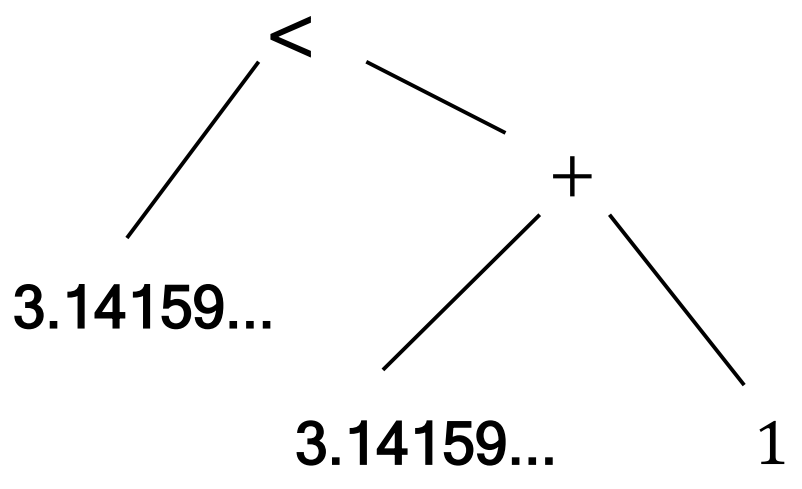

Consequently, sentences *cannot* be enumerated and there are some sentences that *cannot* be obtained by parsing strings. These arrangements exclude known paradoxes from Classical Direct Logic.[iii]

**Note: type theory of Classical Direct Logic is much stronger than constructive type theory with constructive logic[19] because Classical Direct Logic has all of the power of Classical Mathematics.**

---

[i] whose digits are incrementally computable

[ii] Using abstraction on terms. See explanation below.

[iii] Please see historical appendix of this article.

## Mathematics self proves that it is open

Mathematics proves that it is open in the sense that it can prove that its theorems cannot be provably computationally enumerated:[20]

**Theorem** ⊢Mathematics is Open

Proof.[i] Suppose to obtain a contradiction that it is possible to prove closure, *i.e.*, there is a provably computable total procedure Proof such that it is provable that

$$\vdash^{p} \Psi \;\Leftrightarrow\; \exists\,[i{:}\mathbb{N}]\rightarrow\; \text{Proof}[i]{=}p$$

As a consequence of the above, there is a provably total procedure ProvableComputableTotal that enumerates the provably total computable procedures that can be used in the implementation of the following procedure:

Diagonal[i] ≡ (ProvableComputableTotal[i])[i]+1

However,

- ProvableComputableTotal[Diagonal] because Diagonal is implemented using provably computable total procedures.
- ¬ProvableComputableTotal[Diagonal] because Diagonal is a provably computable total procedure that differs from every other provably computable total procedure.

The above contradiction completes the proof.

[Franzén 2004] argued that mathematics is inexhaustible because of inferential undecidability[ii] of mathematical theories. The above theorem that mathematics is open provides another independent argument for the inexhaustibility of mathematics.

## Completeness of inference versus inferential undecidability of closed mathematical theories

A closed mathematical theory is an extension of mathematics whose proofs are computationally enumerable. For example, group theory is obtained by adding the axioms of groups to Classical Direct Logic along with the axiom that theorems of group theory are computationally enumerable.

By definition, if T is a closed theory, there is a total procedure $\text{Proof}_T$ such that $\vdash^{p}_{T} \Psi \Leftrightarrow \exists\,[i{:}\mathbb{N}]\rightarrow \text{Proof}_T[i]{=}p$

[i] This argument appeared in [Church 1934] expressing concern that the argument meant that there is "*no sound basis for supposing that there is such a thing as logic*."

[ii] See section immediately below.

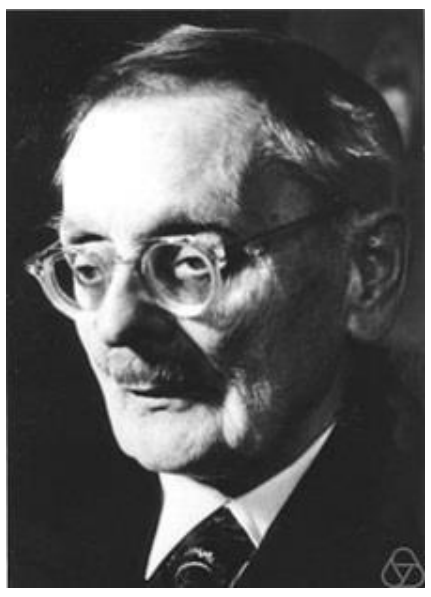

Ernst Zermelo

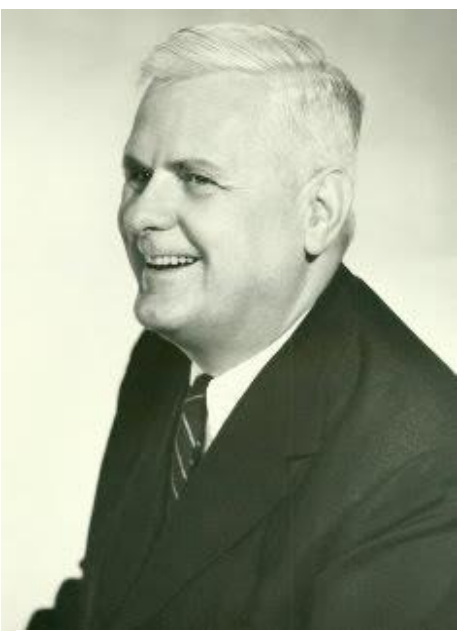

Alonzo Church

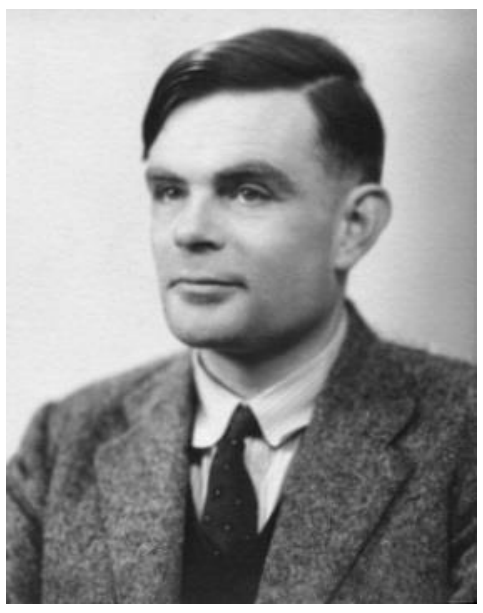

Alan Turing

**Theorem:**[i] If $\mathcal{T}$ is a consistent, closed mathematical theory, there is a proposition ChurchTuring$_{\mathcal{T}}$ such that both of the following hold:[ii]

- $\vdash\nvdash_{\mathcal{T}}$ ChurchTuring$_{\mathcal{T}}$
- $\vdash\nvdash_{\mathcal{T}}\neg$ChurchTuring$_{\mathcal{T}}$

---

[i] First stated in [Gödel 1931].

[ii] Proof [Church 1935, Turing 1936] (This proof is a replacement for the incorrect proof in [Gödel 1931]):

Otherwise, provability in classical logic would be computationally decidable because

$\forall$[p:Program, x:$\mathbb{N}$]→ Halt[p, x] ⇔ ⊢ Halt[p, x]

where Halt[p, x] if and only if program p halts on input x. If such a $\psi_{\text{ChurchTuring}}$ did not exist, then provability could be decided by enumerating theorems until the proposition in question or its negation is encountered.

Note the following important ingredients for the proof of inferential undecidability of a consistent, closed mathematical theory:

- Closure (computational enumerability) of theorems of a mathematical theory to carry through the proof.
- Consistency (nontriviality) to prevent everything from being provable

**Corollary:** There is a proposition Φ of ℕ such that the following hold:[21]

- $\vdash \models_{\mathbb{N}} \Phi$
- $\vdash \nvdash_{\mathbb{N}} \Phi$

Proof.[i] Since ℕ is consistent, one of the following two cases hold:

1. $\models_{\mathbb{N}} \Psi_{\text{ChurchTuring}}$: choose Φ to be $\text{ChurchTuring}_{\mathbb{N}}$
2. $\models_{\mathbb{N}} \neg\Psi_{\text{ChurchTuring}}$: choose Φ to be $\neg\text{ChurchTuring}_{\mathbb{N}}$

Information Invariance[ii] is a *fundamental* technical goal of logic consisting of the following:

1. *Soundness of inference:* information is not increased by inference[iii]
2. *Completeness of inference:* all information that necessarily holds can be inferred

Note that a closed mathematical theory $\mathcal{T}$ is inferentially undecidable[iv] with respect to $\text{ChurchTuring}_{\mathcal{T}}$ does not mean *incompleteness* with respect to the information that can be inferred about theory $\mathcal{T}$ because $\vdash(\nvdash_{\mathcal{T}} \text{ChurchTuring}_{\mathcal{T}}), (\nvdash_{\mathcal{T}} \neg\text{ChurchTuring}_{\mathcal{T}})$.[22]

## Overview

| Contradiction | Outcome |
|---|---|
| Church discovered to his dismay that if theorems of mathematics are postulated to be computationally enumerable, then mathematics is inconsistent. | Theorems of mathematics cannot be computationally enumerated and mathematics is open and inexhaustible. But theorems of a particular theory can be postulated to be computationally enumerable. |

[i] This proof is a replacement for the invalid proof in [Gödel 1931].
[ii] Closely related to conservation laws in physics
[iii] *E.g.* inconsistent information does not infer nonsense.
[iv] sometimes called "incomplete"

| | |
|---|---|
| Using fixed points to construct a "self-referential" sentence for an untyped grammar of mathematical sentences, [Gödel 1931] claimed that mathematics cannot prove its own consistency. However, it is pointed out in this paper that mathematics easily proves its own consistency. | The contradiction can be resolved by using a properly-typed grammar for sentences of mathematics does not allow the use of fixed points to construct "self-referential" sentences.[i] |
| Using fixed points to construct a "self-referential" sentence using an untyped grammar of mathematical sentences, [Gödel 1931] claimed to prove inferential undecidability (sometimes called "incompleteness") for mathematics. However, such "self-referential" sentences lead to inconsistency in mathematics. | [Church 1935, Turing 1936] proved inferential undecidabilty of closed mathematical theories without using fixed points for an untyped grammar of mathematical sentences to construct "self-referential" sentences. |
| In Computer Science, it is important that the Natural Numbers (ℕ) be axiomatized in a way that does not allow integers (e.g. infinite ones) in models of the axioms. However, it is impossible to properly axiomatize ℕ using first-order logic. | Using Classical Direct Logic, ℕ can be axiomatized in such a way that all models are uniquely isomorphic to ℕ [Dedekind 1888, Peano 1889]. Consequently, there are no infinite integers in models of the axioms. |
| In Computer Science, it is important that sets of the Natural Numbers (**Sets**◁ℕ▷) be axiomatized in a way that does not allow countable models. However, it is impossible to properly axiomatize **Sets**◁ℕ▷ using first-order logic. | Using Classical Direct Logic, **Sets**◁ℕ▷ are defined by characteristic functions of types and thus all models are uniquely isomorphic to 𝕊𝕖𝕥𝕤◁ℕ▷. Consequently, its models have no infinite integers or other nonstandard elements. |

[i] Note this does not prevent using fixed points to define recursion in programs.

| | |
|---|---|
| First-order logic is unsuitable as the foundation of mathematics for Computer Science:<br>• Some theorems of ordinary classical mathematics cannot be proved.<br>• Some ordinary theorems useful in Computer Science cannot be proved.<br>• There are undesirable models of mathematical theories (see above). | Classical Direct Logic is suitable as the foundation of mathematics for Computer Science:<br>• All ordinary theorems of classical mathematics can be proved.<br>• All ordinary theorems useful in Computer Science can be proved<br>• There are no undesirable models of mathematical theories. |

## Conclusion

> *The problem is that today some knowledge still feels too dangerous because our times are not so different to Cantor or Boltzmann or Gödel's time. We too feel things we thought were solid being challenged; feel our certainties slipping away. And so, as then, we still desperately want to cling go a belief in certainty. It makes us feel safe. ...*

Are we grown up enough to live with uncertainties or will we repeat the mistakes of the twentieth century and pledge blind allegiance to another certainty. Malone [2007]The world always needs heretics to challenge the prevailing orthodoxies. We are lucky that we can be heretics today without any danger of being burned at the stake. But unfortunately I am an old heretic. Old heretics do not cut much ice. When you hear an old heretic talking, you always say, "Too bad he has lost his marbles."

> *What the world needs is young heretics. I am hoping that one or two of you people in the audience may fill that role.*
> Dyson [2005]

A closed mathematical theory is an extension of mathematics whose proofs are computationally enumerable. For example, group theory is obtained by adding the axioms of groups to Classical Direct Logic along with the axioms that the theorems of group theory are computationally enumerable. If a closed mathematical theory $\mathcal{T}$ is consistent, then it is inferentially undecidable[i] because provability in $\mathcal{T}$ is computationally undecidable [Church 1935 and later Turing 1936].

---

[i] *i.e.* there is a proposition $\Psi$ such that $\nvdash_{\mathcal{T}}\Psi$ *and* $\nvdash_{\mathcal{T}}\neg\Psi$, which is sometimes called "incompleteness"

Information Invariance is a fundamental technical goal of logic consisting of the following:

1. *Soundness of inference*: information is not increased by inference
2. *Completeness of inference*: all information that necessarily holds can be inferred.

That a closed mathematical theory $T$ is inferentially undecidable[i] with respect to Ψ (above) does not mean incompleteness with respect to the information that can be inferred because (by construction) $\vdash(\nvdash_T \Psi), (\nvdash_T \neg\Psi)$.

Computer Science needs a rigorous foundation for all of mathematics that enables computers to carry out all reasoning without human intervention.[23] [Frege 1879] was a good start, but it foundered on the issue of being well-founded. [Russell 1925] attempted basing foundations entirely on types, but foundered on the issue of being expressive enough to carry to some common mathematical reasoning. [Church 1932, 1933] attempted basing foundations entirely on untyped higher-order functions, but foundered because it allowed "self-referential" propositions leading to contradictions [Kleene and Rosser 1935]. Presently, Isabelle [Paulson 1989] and Coq [Coquand and Huet 1986] are founded on types and do not allow theories to reason about themselves. Classical Direct Logic is a foundation for all of mathematical reasoning based on both sets (for well-founded structures) and types (to provide expressibility for concepts including a grammar for propositions) that allows general inference about reasoning.

[Gödel 1931] claimed inferential undecidability[ii] results for Principia Mathematica as the foundation of all of mathematics using a "self-referential" proposition constructed using fixed points exploiting a untyped grammar of mathematical sentences. In opposition to Wittgenstein's correct argument that "self-referential" propositions lead to contradictions in mathematics, Gödel later claimed that his results were for a cut-down first-order theory of Peano numbers. However, first-order logic is not a suitable foundation for Computer Science because of the requirement that computer systems be able to carry out all reasoning without requiring human intervention (including reasoning about their own inference systems). Following [Frege 1879, Russell 1925, and Church 1932-1933], Direct Logic was developed and then investigated "self-referential" propositions with the following results.

- Formalization of Wittgenstein's proof that Gödel's "self-referential" proposition leads to contradiction in mathematics. So the consistency of mathematics had to be rescued against Gödel's "self-referential"

---

[i] sometimes called "incomplete"

[ii] sometimes called "incompleteness"

propositions. The "self-referential" proposition used in results of [Curry 1941] and [Löb 1955] also lead to inconsistency in mathematics. Consequently, mathematics had to be rescued against these "self-referential" propositions as well.

- Self-proof of the consistency of mathematics. Consequently, mathematics had to be rescued against the claim [Gödel 1931] that mathematics cannot prove its own consistency. Also, it became an open problem whether mathematics proves its own consistency, which was resolved by the author discovering an amazing simple proof. [24]

  A solution is to bar "self-referential" propositions using a properly constructed grammar for sentences of mathematics.[25] However, Establishment Philosophers have very reluctant to accept the solution.

  According to [Dawson 2006]:[26]
  - *Gödel's results altered the mathematical landscape, but they did* ***not*** *"produce a debacle".*
  - *There is* ***less*** *controversy today over mathematical foundations than there was* ***before*** *Gödel's work.*

  However, Gödel's results have produced a controversy of a very different kind from the one discussed by Dawson:
  - Gödel's result that mathematics cannot prove its own consistency[i] has been disproved.
  - Consequently, Gödel's results have led to increased controversy over mathematical foundations.

The development of Direct Logic has strengthened the position of working mathematicians as follows:[ii]

- Allowing freedom from the philosophical dogma of the First-Order Thesis
- Providing a usable type theory for all of Mathematics
- Allowing theories to freely reason about theories
- Providing Inconsistency Robust Direct Logic for safely reasoning about theories of practice that are (of necessity) pervasively inconsistent.

[i] Gödel's result was accepted doctrine by Establishment Philosophers for over eight decades

[ii] Of course, Direct Logic must preserve as much previous learning as possible.

**Acknowledgments**
Extensive discussions with Tom Costello, Eric Kao, Ron van der Meyden and other members of the Stanford CS Logic Group helped the development of this paper. Tom suggested that more conventional terminology be used in the formal proof of consistency of mathematics. Martin Davis kindly provide the reference for [Gödel 1933]. Comments by James Lottes, Pat Suppes, Daniel Raggi, Eric Winsberg, John Woods, and Ming Xiong helped improve the presentation. Dan Flickinger suggested including an overview table. Alan Bundy pointed out many crucial places where the presentation needed improvement. John Woods served ably as the senior referee by compiling an excellent synopsis of anonymized conference referee reports for this article. Discussions with Michael Beeson helped improve the section on how mathematics self-proves its own consistency. Correspondence with Monroe Eskew helped clarify the relationship of Classical Direct Logic with first-order logic. Also, Monroe suggested looking at Berry's Paradox.[i] Correspondence with Jack Copeland helped clarify the relationship of the work reported in this article with previous work by Gödel *et. al.* Also, Jack suggested inclusion of the closely related natural deduction proof that mathematics proves its own consistency in addition to the linear proof. Discussions on the FriAM electronic mailing list were very helpful in improving this article.

[i] Please see section on Berry's Paradox in the historical appendix.

## Appendix 1. Notation of Classical Direct Logic

Types and Propositions are defined as follows:

- ***Types***
  - **Boolean**,ℕ[27],**Sentence**,**Proposition**,**Proof**,**Theory**:**Type**
  - If $\sigma_1,\sigma_2$:**Type**, then $\sigma_1 \sqcup \sigma_2$,$[\sigma_1, \sigma_2]$[28],$[\sigma_1] \mapsto \sigma_2$[i],$\sigma_2{}^{\sigma_1}$[ii]:**Type**.
  - If $\sigma$:**Type**, then **Term**$\triangleleft\sigma\triangleright$[29]:**Type**.
  - If $\sigma_1,\sigma_2$:**Type**, $f{:}\sigma_2{}^{\sigma_1}$ and $x{:}\sigma_1$, then $f[x]{:}\sigma_2$.
  - If $\sigma_1,\sigma_2$:**Type**, then $\sigma_1 \sqcup \sigma_2$, $[\sigma_1] \mapsto \sigma_2$, $\sigma_2{}^{\sigma_1}$:**Type**
  - If $\sigma$:**Type**, then **Term**$\triangleleft\sigma\triangleright$:**Type**

- ***Propositions***, *i.e.,* x:**Proposition** ⇔ x constructed by the rules below:
  - If $\sigma$:Type, $\Pi$:Boolean$^{\sigma}$ and $x{:}\sigma$, then $\Pi[x]$:Proposition.[iii]
  - If $\Phi$:Proposition, then $\neg\Phi$:Proposition.
  - If $\Phi,\Psi$:Proposition, then $\Phi\wedge\Psi$, $\Phi\vee\Psi$, $\Phi\Rightarrow\Psi$, $\Phi\Leftrightarrow\Psi$:Proposition.
  - If p:Boolean and $\Phi,\Psi$:Proposition, then (p �True⸽ $\Phi_1$, False⸽ $\Phi_2$):Proposition.[30]
  - If $\sigma_1,\sigma_2$:Type, $x_1{:}\sigma_1$ and $x_2{:}\sigma_2$, then $x_1{=}x_2, x_1{\in}x_2, x_1{\sqsubseteq}x_2, x_1{\subseteq}x_2, x_1{:}x_2$:Proposition.
  - If $\Phi_{1\ \text{to}\ n}$:Proposition, then $(\Phi_1, \ldots, \Phi_k \vdash \Phi_{k+1}, \ldots, \Phi_n)$:Proposition [31]
  - If p:Proof and $\Phi$:Proposition, then $(\overset{p}{\vdash} \Phi)$:Proposition[iv]

---

[i] type of computable procedures from type $\sigma_1$ into $\sigma_2$.

[ii] type of functions from **σ1** into **σ2**

[iii] $\Pi[x] \Leftrightarrow (\Pi[x]=\text{True})$

Note that $\sigma$:Type, $\Pi$:Boolean$^{\sigma}$ means that there are no fixed points for propositions.

[iv] **p** is a proof of $\Phi$

Grammar trees (*i.e.* expressions, terms, and sentences) are defined as follows :

- ***Expressions***, *i.e.,* x:Expression◁σ▷ ⇔ x constructed by the rules below:
  - **True**,**False**:Constant◁Boolean▷ and **0**,**1**:Constant◁ℕ▷.
  - If **σ**:Type and **x**:Constant◁σ▷, then **x**:Expression◁σ▷.
  - If **σ**:Type and **x**:Variable◁σ▷, then **x**:Expression◁σ▷.
  - If **σ**,$\sigma_{1 \text{ to } n}$:Type, $x_{1 \text{ to } n}$:Expression◁$\sigma_{1 \text{ to } n}$▷ and **y**:Expression◁σ▷, then (**Let** {$v_1 \equiv x_1$ , … , $v_n \equiv x_n$}, **y**):Expression◁σ▷ and $v_{1 \text{ to } n}$:Variable◁$\sigma_{1 \text{ to } n}$▷ in **y** and in each $x_{1 \text{ to } n}$.
  - If $e_1$, $e_2$:Expression◁Type▷, then ⌈$e_1$⊔$e_2$⌉,⌈ [$e_1$, $e_2$] ⌉,⌈[$e_1$]↦$e_2$⌉,⌈$e_2{}^{e_1}$⌉:Expression◁Type▷.
  - If $t_1$:Expression◁Boolean▷, $t_2$,$t_3$:Expression◁σ▷, then ⌈$t_1$ �True ⸬ $t_2$, False ⸬ $t_3$⌉:Expression◁σ▷.[32]
  - If $\sigma_1$,$\sigma_2$:Type, **t**:Expression◁$\sigma_2$▷, then ⌈[x:$\sigma_1$]→ t⌉:Expression◁[$\sigma_1$]↦$\sigma_2$▷ and **x**:Variable◁$\sigma_1$▷.[33]
  - If $\sigma_1$,$\sigma_2$:Type, **p**:Expression◁[$\sigma_1$]↦$\sigma_2$▷ and **x**:Expression◁$\sigma_1$▷, then ⌈p▪[x] ⌉:Expression◁$\sigma_2$▷.
  - If **e**:Expression◁σ▷, then ⌈⌊e⌋⌉:Expression◁σ▷.
  - If **e**:Expression◁σ▷ with no free variables and **e** converges, then ⌊e⌋:σ.

- ***Terms***, *i.e.,* x:Term◁σ▷ ⇔ x constructed by the rules below:
  - If **σ**:Type and **x**:Constant◁σ▷, then **x**:Term◁σ▷.
  - If **σ**:Type and **x**:Variable◁σ▷, then **x**:Term◁σ▷.
  - If $t_1$, $t_2$:Term◁Type▷, then ⌈$t_1$⊔$t_2$⌉,⌈[$t_1$, $t_2$] ⌉,⌈[$t_1$]↦$t_2$⌉,⌈$t_2{}^{t_1}$⌉:Term◁Type▷.
  - If **σ**:Type, $t_1$:Term◁Boolean▷, $t_2$,$t_3$:Term◁σ▷, then ⌈$t_1$ �True ⸬ $t_2$, False ⸬ $t_3$⌉:Term◁σ▷.
  - If $\sigma_1$,$\sigma_2$:Type, **f**:Term◁$\sigma_2{}^{\sigma_1}$▷ and **t**:Term◁$\sigma_1$▷, then ⌈f[t] ⌉:Term◁$\sigma_2$▷.
  - If $\sigma_1$,$\sigma_2$:Type and **t**:Term◁$\sigma_2$▷, then ⌈[x:$\sigma_1$]→ t⌉:Term◁$\sigma_2{}^{\sigma_1}$▷ and **x**:Variable◁σ1▷ in **t**.
  - If **σ**:Type and **t**:Term◁σ▷, then ⌈⌊t⌋⌉:Term◁σ▷.
  - If **σ**:Type, **e**:Expression◁σ▷ with no free variables and **e** converges, then **e**:Constant◁σ▷.
  - If **σ**:Type and **t**:Term◁σ▷ with no free variables, then ⌊t⌋:σ.

- ***Sentences**, i.e.,* x:**Sentence** ⇔ x constructed by the rules below:
  - If $s_1$:**Sentence** then, ⌈¬$s_1$⌉:**Sentence**.
  - If $s_1$,$s_2$:**Sentence** then ⌈$s_1 \wedge s_2$⌉,⌈$s_1 \vee s_2$⌉,⌈$s_1 \Rightarrow s_2$⌉,⌈$s_1 \Leftrightarrow s_2$⌉:**Sentence**.
  - If σ:**Type**, $t_1$:**Term**◁**Boolean**$^{\sigma}$▷ and $t_2$:**Term**◁σ▷, then ⌈$t_1[t_2]$ ⌉:**Sentence**
  - If t:**Term**◁**Boolean**▷ and $s_1$,$s_2$:**Sentence**, then ⌈t �**True** ⦂ $s_1$, **False** ⦂ $s_2$⌉:**Sentence**.[34]
  - If $\sigma_1$,$\sigma_2$:**Type**, $t_1$:**Term**◁$\sigma_1$▷ and $t_2$:**Term**◁$\sigma_2$▷, then ⌈$t_1$=$t_2$⌉,⌈$t_1 \in t_2$⌉,⌈$t_1 \sqsubseteq t_2$⌉,⌈$t_1 \subseteq t_2$⌉,⌈$t_1$:$t_2$⌉:**Sentence**.
  - If σ:**Type** and s:**Sentence**, then ⌈∀[x:σ]→ s⌉,⌈∃[x:σ]→ s⌉:**Sentence** and x:**Variable**◁σ▷ in **s.**
  - If $s_{1 \text{ to } n}$:**Sentence**, then ⌈$s_1, \ldots, s_k \vdash s_{k+1}, \ldots, s_n$⌉:**Sentence**
  - If p:**Term**◁**Proof**▷ and s:**Sentence**, then ⌈ $\overset{p}{\vdash}$ s⌉:**Sentence**
  - If s:**Sentence**, then ⌈ ⌊s⌋⌉:**Sentence**.
  - If s:**Sentence** with no free variables, then ⌊s⌋:**Proposition**.

## Foundations with both types and sets

> *Everyone is free to elaborate* [their] *own foundations. All that is required of* [a] *Foundation of Mathematics is that its discussion embody absolute rigor, transparency, philosophical coherence, and addresses fundamental methodological issues.*
> [Nielsen 2014]

Classical Direct Logic develops foundations for mathematics using *both*[i] types[ii] *and* sets[iii] encompassing all of standard mathematics including the integers, reals, analysis, geometry, *etc.*[35]

Combining types and sets as the foundation has the advantage of using the strengths of each without the limitations of trying to use just one because each can be used to make up for the limitations of the other. The key idea is

---

[i] Past attempts to reduce mathematics to logic alone, to sets alone, or to types alone have not be very successful.

[ii] According to [Scott 1967]: *"*there is only one satisfactory way of avoiding the paradoxes: namely, the use of some form of the *theory of types*... the best way to regard Zermelo's theory is as a simplification and extension of Russell's ...*simple* theory of types. Now Russell made his types *explicit* in his notation and Zermelo left them *implicit*. It is a mistake to leave something so important invisible...*"*

[iii] According to [Scott 1967]: *"*As long as an idealistic manner of speaking about abstract objects is popular in mathematics, people will speak about collections of objects, and then collections of collections of ... of collections. In other words *set theory is inevitable."* [emphasis in original]

compositionality, *i.e.*, composing new entities from others. Types can be composed from other types and sets can be composed from other sets.

Functions, graphs, and lists are fundamental to the mathematical foundations of Computer Science. **SetFunctions**◁**σ**▷ (type of set functions based on type **σ**) that can be defined inductively as follows:

**SetFunctionsOfOrder**◁**σ**▷[1] ≡ $\boldsymbol{\sigma}^{\boldsymbol{\sigma}}$

**SetFunctionsOfOrder**◁**σ**▷[n+1] ≡

$$(\boldsymbol{\sigma}\sqcup\textbf{SetFunctionsOfOrder}\triangleleft\boldsymbol{\sigma}\triangleright[n])^{\boldsymbol{\sigma}\sqcup\textbf{SetFunctionsOfOrder}\triangleleft\boldsymbol{\sigma}\triangleright[n]}$$

Furthermore the process of constructing orders of **SetFunctionsOfOrder**◁**σ**▷is exhaustive for SetFunctions◁**σ**▷:[i]

$$\textbf{SetFunctions}\triangleleft\boldsymbol{\sigma}\triangleright \equiv \coprod_{i:\mathbb{N}} \textbf{SetFunctionsOfOrder}\triangleleft\boldsymbol{\sigma}\triangleright[i]$$

Sets (along with lists) provide a convenient way to collect together elements.[36] For example, sets (of sets of sets of ...) of **σ** can be axiomatized as follows:[ii]

∀[s:**Sets**◁**σ**▷]→ ∃[f:**SetFunctions**◁**σ**▷]→ CharacteristicFunction[f, s]

where ∀[s:**Sets**◁**σ**▷, f:$\textbf{Boolean}^{\textbf{SetFunctions}\triangleleft\boldsymbol{\sigma}\triangleright}$]→

CharacteristicFunction[f, s]

⇔ ∀[e:**σ**⊔**Sets**◁**σ**▷]→ e∈s ⇔ f[e]=True

*i.e.* every set of type **Sets**◁**σ**▷ is defined by a characteristic function of **SetFunctions**◁**σ**▷

Note that there is no set corresponding to the *type* **Sets**◁ℕ▷which is an example of how types extend the capabilities of sets.[37]

Although **Sets**◁ℕ▷ are well-founded[38], in general sets in Direct Logic are not well-founded. For example, consider the following definition:

InfinitelyDeep.[ ] ≡ {**postpone** InfinitelyDeep.[ ]}[iii]

Consequently, InfinitelyDeep.[ ]∈InfinitelyDeep.[ ].

---

[i] The closure property below is used to guarantee that there is just one model of **SetFunctions**◁ℕ▷ up to isomorphism using a unique isomorphism.

[ii] Of course, the higher cardinals are left out of these foundations. On the other hand, Computer Science doesn't need higher cardinals in its mathematical foundations.

[iii] InfinitelyDeep.[ ] = {{{{{...}}}}}

## Natural Numbers, Real Numbers, and their Sets are Unique up to Isomorphism[i]

The following question arises: What mathematics have been captured in the above foundations?

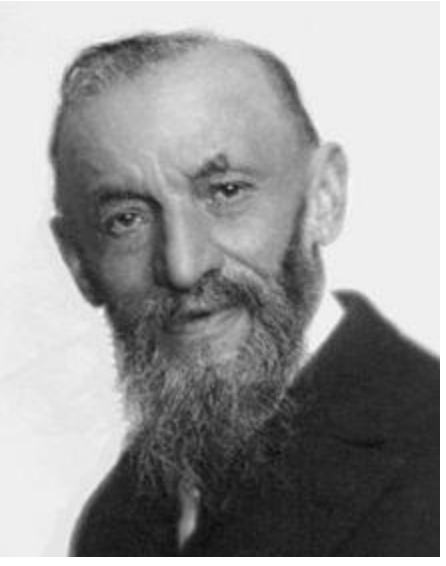

Giuseppe Peano

**Theorem[ii] (Categoricity of $\mathbb{N}$):**[39]

$\forall[\mathbb{M}{:}\mathbf{Model}\triangleleft\mathbb{N}\triangleright]\to \mathbb{M}\approx\mathbb{N}$, *i.e.*, models of the natural numbers $\mathbb{N}$ are isomorphic by a unique isomorphism.[iii]

The following strong induction axiom[40] can be used to characterize the natural numbers ($\mathbb{N}$[41]) up to isomorphism with a unique isomorphism:

$\forall[P{:}\mathbf{Boolean}^{\mathbb{N}}]\to\ \text{Inductive}[P] \Rightarrow \forall[i{:}\mathbb{N}]\to P[i]$

*where* $\forall[P{:}\mathbf{Boolean}^{\mathbb{N}}]\to \text{Inductive}[P]$

$\Leftrightarrow (P[0] \wedge \forall[i{:}\mathbb{N}]\to P[i] \Rightarrow P[i+1])$[42]

**Theorem[iv] (Categoricity of $\mathbb{R}$):**[43]

$\forall[\mathbb{M}{:}\mathbf{Model}\triangleleft\mathbb{R}\triangleright]\to \mathbb{M}\approx\mathbb{R}$, *i.e.*, models of the real numbers $\mathbb{R}$ are isomorphic by a unique isomorphism.[v]

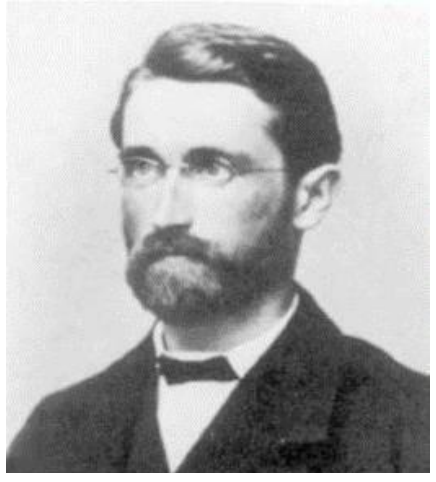

Richard Dedekind

The following can be used to characterize the real numbers ($\mathbb{R}$[44]) up to isomorphism with a unique isomorphism:

$\forall[S{:}\mathbf{Set}\triangleleft\mathbb{R}\triangleright]\to\ S\neq\{\,\} \wedge \text{Bounded}[S] \Rightarrow \text{HasLeastUpperBound}[S]$

*where*

$\text{Bounded}[S{:}\mathbf{Set}\triangleleft\mathbb{R}\triangleright] \Leftrightarrow \exists[b{:}\mathbb{R}]\to \text{UpperBound}[b, S]$

$\text{UpperBound}[b{:}\mathbb{R}, S{:}\mathbf{Set}\triangleleft\mathbb{R}\triangleright] \Leftrightarrow\ b\in S \wedge \forall[x\in S]\to x\leqq b$

$\text{HasLeastUpperBound}[S{:}\mathbf{Set}\triangleleft\mathbb{R}\triangleright]] \Leftrightarrow \exists[b{:}\mathbb{R}]\to \text{LeastUpperBound}[b, S]$

$\text{LeastUpperBound}[b{:}\mathbb{R}, S{:}\mathbf{Set}\triangleleft\mathbb{R}\triangleright]$

$\Leftrightarrow\ \text{UpperBound}[b,S] \wedge \forall[x\in S]\to \text{UpperBound}[x,S] \Rightarrow x\leqq b$

---

[i] and the isomorphism is unique

[ii] [Dedekind 1888, Peano 1889]

[iii] Consequently, the type of natural numbers $\mathbb{N}$ is unique up to isomorphism and the type of reals $\mathbb{R}$ is unique up to isomorphism.

[iv] [Dedekind 1888]

[v] Consequently, the type of natural numbers $\mathbb{N}$ is unique up to isomorphism and is a subtype of reals $\mathbb{R}$ that is unique up to isomorphism.

**Theorem (Categoricity of Sets◁ℕ⊔ℝ▷):**[45]

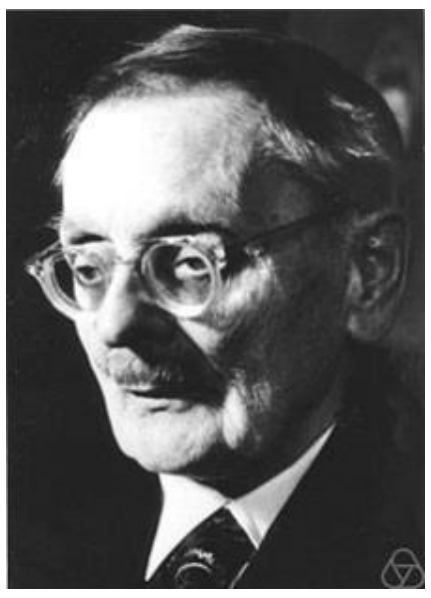

Ernst Zermelo

$\forall[\mathbb{M}{:}\text{Model}\triangleleft\textbf{Sets}\triangleleft\mathbb{N}\sqcup\mathbb{R}\triangleright\triangleright]\rightarrow\ \mathbb{M}\approx\mathbb{S}\mathbb{e}\mathbb{t}\mathbb{s}\triangleleft\mathbb{N}\sqcup\mathbb{R}\triangleright$, *i.e.*, models of **Sets**◁ℕ⊔ℝ▷ are isomorphic by a unique isomorphism.[i]

**Sets**◁ℕ⊔ℝ▷ (which is a fundamental type of mathematics) is exactly characterized axiomatically, which is what is required for Computer Science.

Proof: By above, $\forall[\mathbb{M}{:}\textbf{Model}\triangleleft\mathbb{N}\triangleright]\rightarrow\mathbb{M}\approx\mathbb{N}$, *i.e.*, models of ℕ are isomorphic by a unique isomorphism. Unique isomorphism of higher order sets can be proved using induction from the following closure property for SetFunctions (see above):

$$\textbf{SetFunctions}\triangleleft\mathbb{N}\sqcup\mathbb{R}\triangleright\equiv\coprod_{i:\mathbb{N}}\textbf{SetFunctionsOfOrder}\triangleleft\mathbb{N}\sqcup\mathbb{R}\triangleright[i]$$

Unique isomorphism for **SetFunctions**◁ℕ⊔ℝ▷ can be extended to **Sets**◁ℕ⊔ℝ▷ because every set in Sets◁ℕ⊔ℝ▷ is defined by a characteristic function of **SetFunctions**◁ℕ⊔ℝ▷. (See above)

Classical Direct Logic is much stronger than first-order axiomatizations of set theory.[46]

**Theorem (Set Theory Model Soundness):** $(\vdash_{\textbf{Sets}\triangleleft\mathbb{N}\triangleright}\Psi)\Rightarrow\ \vDash_{\mathbb{S}\mathbb{e}\mathbb{t}\mathbb{s}\triangleleft\mathbb{N}\triangleright}\Psi$

Proof: Suppose $\vdash_{\textbf{Sets}\triangleleft\mathbb{N}\triangleright}\Psi$. The conclusion immediately follows because the axioms for the theory **Sets**◁ℕ▷ hold in the model 𝕊𝕖𝕥𝕤◁ℕ▷.

[i] Consequently, the type of natural numbers ℕ is unique up to isomorphism and the type of reals ℝ is unique up to isomorphism.

## Appendix 2. Historical Background

> *The powerful (try to) insist that their statements are literal depictions of a single reality. 'It really is that way', they tell us. 'There is no alternative.' But those on the receiving end of such homilies learn to read them allegorically, these are techniques used by subordinates to read through the words of the powerful to the concealed realities that have produced them.* Law [2004]

### *Gödel was certain*

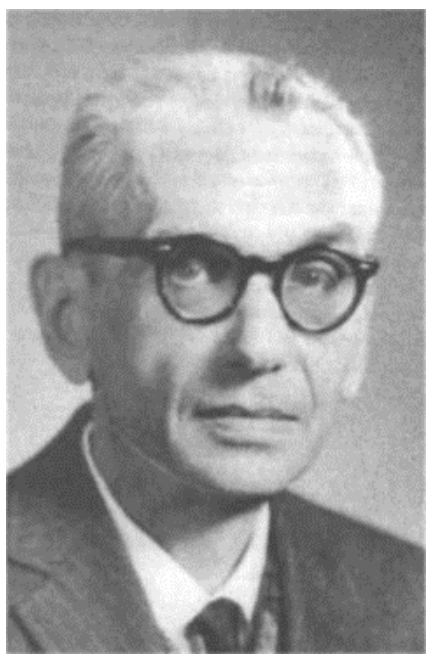

Kurt Gödel

> *"Certainty" is far from being a sign of success; it is only a symptom of lack of imagination and conceptual poverty. It produces smug satisfaction and prevents the growth of knowledge.* [Lakatos 1976]

Paul Cohen [2006] wrote as follows of his interaction with Gödel:

> *His* [Gödel's] *main interest seemed to lie in discussing the "truth" or "falsity" of these questions, not merely in their undecidability. He struck me as having an almost unshakable belief in this "realist" position, which I found difficult to share. His ideas were grounded in a deep philosophical belief as to what the human mind could achieve. I greatly admired this faith in the power and beauty of Western Culture, as he put it, and would have liked to understand more deeply what were the sources of his strongly held beliefs. Through our discussions, I came closer to his point of view, although I never shared completely his "realist" point of view, that all questions of Set Theory were in the final analysis, either true or false.*

Chaitin [2007] presented the following analysis:

> Gödel's proof of inferential undecidability [incompleteness] *was too superficial. It didn't get at the real heart of what was going on. It was more tantalizing than anything else. It was not a good reason for something so devastating and fundamental. It was too clever by half. It was too superficial.* [It was based on the clever construction] *"I'm unprovable." So what? This doesn't give any insight how serious the problem is.*

After Church[1935] and Turing[1936] proved inferential undecidabilty of closed mathematical theories using computational undecidablity[i], Gödel claimed more generality and that his results applied to all consistent mathematical systems that incorporate Peano axioms. However, when he learned of Wittgenstein's devastating proof of inconsistency,[47] Gödel retreated to claiming that his results applied to the very weak system of first-order Peano.[48] The upshot is that Gödel never acknowledged that his "self-referential" proposition[i] implies inconsistency in mathematics. See further discussion below in this article.

[i] See proof of inferential undecidablity of closed mathematical theories elsewhere in this article.

Also, the ultimate criteria for correctness in the theory of natural numbers is provability using strong induction [Dedekind 1888, Peano 1889]. In this sense, Wittgenstein was correct in his identification of "truth" with provability. On the other hand, Gödel obfuscated the important identification of provability as the touchstone of ultimate correctness in mathematics.

von Neumann [1961] had a very different view from Gödel:

> *It is* ***not*** *necessarily true that the mathematical method is something absolute, which was revealed from on high, or which somehow, after we got hold of it, was evidently right and has stayed evidently right ever since.*

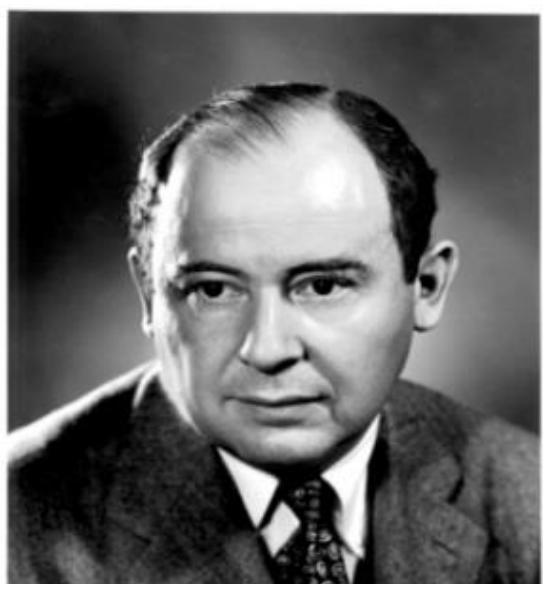

John von Neumann

## Limitations of first-order logic

> *By this it appears how necessary it is for nay man that aspires to true knowledge to examine the definitions of former authors; and either to correct them, where they are negligently set down, or to make them himself. For the errors of definitions multiply themselves, according as the reckoning proceeds, and lead men into absurdities, which at last they see, but cannot avoid, without reckoning anew from the beginning; in which lies the foundation of their errors...*
> [Hobbes *Leviathan*, Chapter 4][49]

It is very important not to confuse Mathematics with first-order logic, which was invented by philosophers for their own purposes. It turns out that first-order logic is amazing weak. For example, first-order logic is incapable of characterizing even the Peano numbers, *i.e.*, there are infinite integers in models of every first-order axiomatization of the Peano numbers. Furthermore, there are infinitesimal real numbers in models of every first-order axiomatization of the real numbers.[i] Of course, infinite integers and infinitesimal reals are monsters that must be banned from the mathematical foundations of Computer Science.

---

[i] constructed using fixed points exploiting an untyped grammar of mathematics

However, some philosophers have found first-order logic to be useful for their careers because it is weak enough that they can prove theorems about first-order axiomatizations whereas they cannot prove such theorems about stronger practical systems, *e.g.*, Classical Direct Logic. For example, there is a famous theorem that first-order set theory is too weak to decide ContinuumHypothesis[50], *i.e.*, $\nvdash_{\text{FirstOrderSetAxioms}}$ContinuumHypothesis and $\nvdash_{\text{FirstOrderSetAxioms}}\neg$ContinuumHypothesis.[51] However, ContinuumHypothesis is still an open problem in Mathematics. That ContinuumHypothesis is an open problem is not so important for Computer Science because for subsets of reals of that are computable[ii], the ComputationalContinuumTheorem[iii] holds.[52]

Zermelo considered the First-Order Thesis to be a mathematical “hoax” because it necessarily allowed unintended models of axioms.[53]

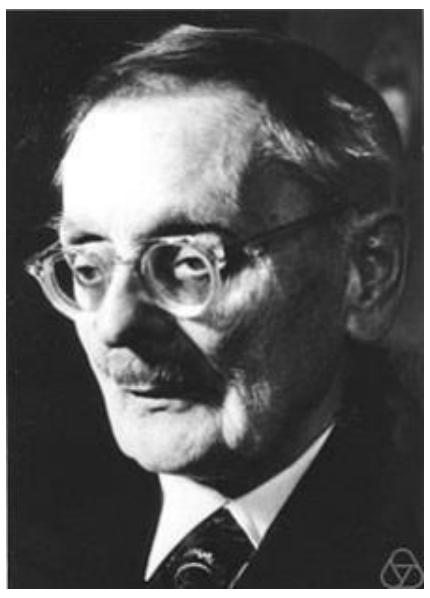

Ernst Zermelo

[i] Likewise, first-order set theory (e.g. ZFC) is very weak. See discussion immediately below.

[ii] A real number is computable if an only if its digits are computable.

[iii] $\mathbb{R}$ has no subset of computable reals whose cardinality is strictly between $\mathbb{N}$ and $\mathbb{R}$.

[Barwise 1985] critiqued the First-Order Thesis[i] as follows:

> *The reasons for the widespread, often uncritical acceptance of the first-order thesis are numerous. The first-order thesis ... confuses the subject matter of logic with one of its tools. First-order language is just an artificial language structured to help investigate logic, much as a telescope is a tool constructed to help study heavenly bodies. From the perspective of the mathematics in the street, the first-order thesis is like the claim that astronomy is the study of the telescope.*[54]

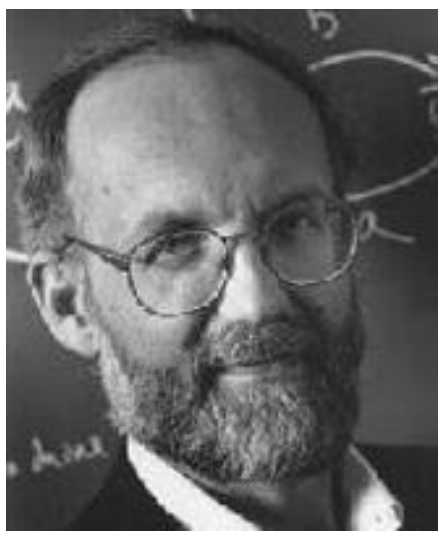

Jon Barwise

Computer Science is making increasing use of Model Analysis[ii] in the sense of analyzing relationships among the following:

- concurrent programs and their Actor Model denotations
- axiom systems and their models

Having infinite integers and infinitesimal reals in models of axioms can cause problems in practical Model Analysis because a computer system can easily prove that there are no infinite integers and no infinitesimal reals. Consequently, infinite integers and infinitesimal reals are modeling monsters. Fortunately, these modeling monsters do not exist in Classical Direct Logic.

The cut-down-first-order theory *FirstOrderPeano*[iii] is too limited for Computer Science because of the following:

- $(\vdash_{\mathit{FirstOrderPeano}} \Psi) \Rightarrow \vdash_{\mathit{Peano}} \Psi$
- There are some $\Psi_0$ that are important in Computer Science (see immediately below) such that:
  - $\vdash_{\mathit{Peano}} \Psi_0$
  - $\nvdash_{\mathit{FirstOrderPeano}} \Psi_0$

In Computer Science, it is important that the Natural Numbers ($\mathbb{N}$) be axiomatized in a way that does not allow non-numbers (*e.g.* infinite ones) in models of the axioms. Unfortunately, every consistent first-order axiomatization of $\mathbb{N}$ has a model with an infinite integer:

---

[i] The "First-Order Thesis" is that mathematical foundations should be restricted to first-order logic.

[ii] a restricted form of Model Checking in which the properties checked are limited to those that can be expressed in Linear-time Temporal Logic has been studied [Clarke, Emerson, Sifakis, *etc.* ACM 2007 Turing Award].

[iii] with cut-down first-order Peano axioms

Theorem: If ℕ is a model of a first-order axiomatization $\mathcal{T}$, then $\mathcal{T}$ has a model 𝕄 with an infinite integer.

Proof: The model 𝕄 is constructed as an extension of ℕ by adding a new element ∞ with the following atomic relationships:

$$\{\neg\infty<\infty\} \cup (\text{Elementwise}[\,[m]\to m<\infty])[\mathbb{N}]^{i}$$

It can be shown that 𝕄 is a model of $\mathcal{T}$ with an infinite integer ∞.

The infinite integer ∞ is a monster that must be banned from the mathematical foundations of Computer Science.

A similar result holds for the standard theory ℝ of real numbers [Dedekind 1888] compared to a cut-down, first-order theory[55], which has models with infinitesimals:

Theorem: If ℝ is a model of a first-order axiomatization $\mathcal{T}$, then $\mathcal{T}$ has a model 𝕄 with an infinitesimal.

Proof: The model 𝕄 is constructed as an extension of ℝ by adding a new element ∞ with the following atomic relationships:

$$\{\neg\infty<\infty\} \cup (\text{Elementwise}[\,[m]\to m<\infty])[\mathbb{N}]^{ii}$$

Defining ε to be $\frac{1}{\infty}$, it follows that $\forall[r{:}\mathbb{R}]\to 0<\varepsilon<\frac{1}{r}$. It can be shown that 𝕄 is a model of $\mathcal{T}$ with an infinitesimal ε, which is a monster that must be banned from the mathematical foundations of Computer Science.

On the other hand, since it is not limited to first-order logic, Classical Direct Logic characterizes structures such as natural numbers and real numbers up to isomorphism.[iii]

---

[i] Elementwise[f] ≡ [s]→ {f[x] | x∈s}

[ii] Elementwise[f] ≡ [s]→ {f[x] | x∈s}

[iii] proving that software developers and computer systems are using the same structures

Of greater practical import, that a computer provides service *i.e.* ∃[i:ℕ]→ ResponseBefore[i] cannot be proved in a first-order theory.

Proof: In order to obtain a contradiction, suppose that it is possible to prove the theorem that computer server provides service[i] in a first-order theory $\mathcal{T}$.
Therefore the following infinite set of propositions is inconsistent:[ii]

(Elementwise[ [i]→ ¬ResponseBefore[i] ])[ℕ].

By the compactness theorem of first-order logic, it follows that there is finite subset of the set of propositions that is inconsistent. But this is a contradiction, because all the finite subsets are consistent since the amount of time before a server responds is unbounded, i.e.,

(∄[i:ℕ]→ $\vdash_{\mathcal{T}}$ ResponseBefore[i]).

The above examples illustrate the following fundamental limitation of first-order theories:

In a first-order theory $\mathcal{T}$, it is impossible to have both of the following for a predicate P:

- ∄[i:ℕ]→ $\vdash_{\mathcal{T}}$ P[i][iii]
- $\vdash_{\mathcal{T}}$ ∃[i:ℕ]→ P[i]

Proof: Suppose that it is possible for both of the above to hold in a first-order theory $\mathcal{T}$. Therefore the following infinite set of propositions is inconsistent:[56]

(Elementwise[ [m]→ ¬P[m] ])[ℕ]

By the compactness theorem of first-order logic, it follows that there is finite subset of the set of propositions that is inconsistent. But this is a contradiction, because all the finite subsets are consistent.

As a foundation of mathematics for Computer Science, Classical Direct Logic provides categorical[57] numbers (integer and real), sets, lists, trees, graphs, etc. which can be used in arbitrary mathematical theories including theories for categories, large cardinals, first-order axiomatizations, etc. These various theories might have "monsters" of various kinds. However, these monsters are not imported into the foundations of Computer Science.

---

[i] ∃[i:ℕ]→ ResponseBefore[i]

[ii] *i.e.* in classical notation: {¬ResponseBefore[i] | i:ℕ}

[iii] *i.e.* ∀[i:ℕ]→ $\nvdash_{\mathcal{T}}$ P[i]

Computer Science needs *stronger* systems than provided by first-order logic in order to weed out unwanted models. In this regard, Computer Science doesn't have a problem computing with "infinite" objects (*i.e.* Actors) such as $\pi$ and uncountable sets such as the real numbers $\mathbb{R}$.

Of course some problems are theoretically not computable. However, even in these cases, it is often possible to compute approximations and cases of practical interest.[i]

The mathematical foundation of Computer Science is very different from the general philosophy of mathematics in which infinite integers and infinitesimal reals may be of some interest. Of course, it is always possible to have special theories with infinite integers, infinitesimal reals, unicorns, *etc*.

## Provability Logic

One kind of Provability Logic (called $\mathcal{PL}$) is a cut-down theory of deduction that has been used to investigate provability predicates for languages that allow "self-referential" propositions [Verbrugge 2010].

In Direct Logic, fixed points on propositions do not exist and consequently Gödel's proofs are not valid. Even in the closed theory $\mathbb{N}$, a "self-referential" sentence cannot be constructed for the following reason:

Let $\text{SentenceFromStringWithIndex}_{\mathbb{N}}$ be a procedure that enumerates sentences that can be produced by parsing mathematical strings of the closed theory $\mathbb{N}$ and $\text{IndexOfSentenceFromString}_{\mathbb{N}}$ be a procedure that returns the index of a sentence that can be obtained by parsing a mathematical string.

$$\text{Gödel:Proposition} \equiv \lfloor \text{SentenceFromStringWithIndex}_{\mathbb{N}}{}_{\blacksquare}[\text{Fix}[\text{Diagonal}]]\rfloor$$

$$\quad \textit{where } \text{Diagonal}_{\blacksquare}[\text{i}:\mathbb{N}]:\mathbb{N} \equiv$$

$$\qquad \text{IndexOfSentenceFromString}_{\mathbb{N}}$$

$$\qquad\quad {}_{\blacksquare}[\ulcorner \nvdash_{\mathbb{N}} \lfloor \text{SentenceFromStringWithIndex}_{\mathbb{N}}{}_{\blacksquare}[\text{i}]\rfloor \urcorner]$$

The fixed point operator Fix cannot be defined using the *strictly typed* lambda calculus because of the following:

$$\textbf{NtoN} \equiv [\mathbb{N}] \mapsto \mathbb{N}$$

$$\text{Helper}_{\blacksquare}[\text{f}:\textbf{NtoN}]:\textbf{NtoN} \equiv [\text{x}:([?] \mapsto \textbf{NtoN}])] \to \text{f}_{\blacksquare}[\text{x}_{\blacksquare}[\text{x}]]$$

$$\text{Fix}_{\blacksquare}[\text{f}:\textbf{NtoN}]:\textbf{NtoN} \equiv (\text{Helper}_{\blacksquare}[\text{f}])_{\blacksquare}[\text{Helper}_{\blacksquare}[\text{f}]]$$

The missing strict type $\boxed{?}$ does not exist.

Because it is first-order, $\mathcal{PL}$ is very weak; even for proving theorems about integers. Also, $\mathcal{PL}$ makes the assumption that there are only countably many

---

[i] *e.g.* see Terminator [Knies 2006], which practically solves the halting problem for device drivers

propositions[58] and that for every proposition $\Phi$, there is an integer $\ulcorner\Phi\urcorner_{\mathcal{PL}}$ such that $\Phi \Leftrightarrow \llcorner\ulcorner\Phi\urcorner_{\mathcal{PL}}\lrcorner_{\mathcal{PL}}$.

In formulating his results, [Löb 1955] proposed the following provability conditions that became the basis of Provability Logic:[i]

1. $(\vdash_{\mathcal{PL}} \Phi) \Rightarrow \vdash_{\mathcal{PL}} \vdash_{\mathcal{PL}} \Phi$
2. $\vdash_{\mathcal{PL}} ((\vdash_{\mathcal{PL}} (\Phi \Rightarrow \Psi)) \Rightarrow ((\vdash_{\mathcal{PL}} \Phi) \Rightarrow \vdash_{\mathcal{PL}} \Psi)))$
3. $\vdash_{\mathcal{PL}} ((\vdash_{\mathcal{PL}} \Phi) \Rightarrow \vdash_{\mathcal{PL}} \vdash_{\mathcal{PL}} \Phi)$

Using "self-referential" propositions, [Löb 1955] proved the following:[59]

$$(\vdash_{\mathcal{PL}} ((\vdash_{\mathcal{PL}} \Phi) \Rightarrow \Phi)) \Rightarrow \vdash_{\mathcal{PL}} \Phi.$$

However, $\mathcal{PL}$ is a very weak theory of deduction. For example, the principle of natural deduction below called "Soundness" in Direct Logic that allows theorems to be used in subarguments is not allowed in $\mathcal{PL}$:[60]

$$(\vdash \Phi) \Rightarrow \Phi$$

Note that the rule of Soundness [*i.e.* $(\vdash \Phi) \Rightarrow \Phi$] does not involve any coding of propositions as integers. It is highly desirable for computer systems to be able to reason about the mathematical foundations of Classical Direct Logic using Classical Direct Logic. Unlike $\mathcal{PL}$, Classical Direct Logic does not require complex circumlocutions (involving coding into integers) that obscure what is going on.

In summary, Provability Logic (although a useful historical development step) is too cumbersome and fragile to serve in the mathematical foundation of Computer Science.

---

[i] His formulation actually used the convoluted coding of propositions into integers.

## Inadequacies of Tarskian Set Models

Tarskian Set Models[61] are inadequate for foundations of Computer Science for they are inadequate to characterize direct inference used by systems to reason about their own inference capabilities.[i]

**But the most fundamental limitation of Tarskian Set Models is that large information theories of practice have *no* models because they are pervasively inconsistent.**

## Church's Paradox

> *in the case of any system of symbolic logic, the set of all provable theorems is* [computationally] *enumerable... any system of symbolic logic not hopelessly inadequate ... would contain the formal theorem that this same system ... was either insufficient* [theorems are not computationally enumerable] *or over-sufficient* [that theorems are computationally enumerable means that the system is inconsistent]...
>
> *This, of course, is a deplorable state of affairs...*
>
> *Indeed, if there is no formalization of logic as a whole, then there is no exact description of what logic is, for it in the very nature of an exact description that it implies a formalization. And if there no exact description of logic, then there is no sound basis for supposing that there is such a thing as logic.*
>
> [Church 1934][62]

[Church 1932, 1933] attempted basing foundations entirely on untyped higher-order functions, but foundered because contradictions emerged because

1. His system allowed "self-referential" propositions [Kleene and Rosser 1935]
2. He believed that theorems must be computationally enumerable.

Our proposal is to address the above issues as follows:

1. Not providing for the construction of "self-referential" propositions in mathematics
2. Mathematics self proves that it is "open" in the sense that theorems are not computationally enumerable (*i.e.* not "closed").[ii]

---

[i] *E.g.* the theorems in this article.

[ii] In other words, the paradox that concerned Church (because he thought that it could mean the demise of formal mathematical logic) has been transformed into fundamental theorem of foundations!

## Curry and Löb Paradoxes

An example of a "self-referential" proposition is *"This proposition is not provable"* that was used by in [Gödel 1931].[63] Unfortunately, allowing construction of "self-referential" propositions[i] results in contradictions [Wittgenstein 1937].[64]

For example the following paradoxes prove *every* proposition using "self-referential" propositions:[65]

- ***Curry's Paradox*** **[**Curry 1941**]:** Suppose that Ψ:Proposition. $Curry_{\Psi}$:Proposition ≡[ii] Fix[$f_{\Psi}$ ]

  *where* $f_{\Psi}$[Φ:Proposition]:Proposition ≡ Φ ⊦ Ψ

  1) $Curry_{\Psi}$ ⇔ ($Curry_{\Psi}$ ⊦ Ψ)
  2) ⊦ ($Curry_{\Psi}$ ⊦ $Curry_{\Psi}$)  // *idempotency*
  3) ⊦ ($Curry_{\Psi}$ ⊦ ($Curry_{\Psi}$ ⊦ Ψ))  // *substituting* 1) *into* 2)
  4) ⊦ ($Curry_{\Psi}$ ⊦ Ψ)  // *contraction*
  5) ⊦ $Curry_{\Psi}$  // *substituting* 1) *into* 4)
  6) ⊦ Ψ  // *chaining* 4) *and* 5)

- ***Löb's Paradox*** **[**Löb 1955**]:**[66] Suppose that Ψ:Proposition. $Löb_{\Psi}$:Proposition ≡[iii] Fix[$f_{\Psi}$ ]

  *where* $f_{\Psi}$[Φ:Proposition]:Proposition ≡ (⊦ Φ) ⊦ Ψ

  1) $Löb_{\Psi}$ ⇔ ((⊦ $Löb_{\Psi}$) ⊦ Ψ)
  2) ⊦ ((⊦ $Löb_{\Psi}$) ⊦ $Löb_{\Psi}$)  // *soundness*
  3) ⊦ ((⊦ $Löb_{\Psi}$) ⊦ ((⊦ $Löb_{\Psi}$) ⊦ Ψ))  // *substituting* 1) *into* 2)
  4) ⊦ ((⊦ $Löb_{\Psi}$) ⊦ Ψ)  // *contraction*
  5) ⊦ $Löb_{\Psi}$  // *substituting* 1) *into* 4)
  6) Ψ  // *chaining* 4) *and* 5)

Of course, it is completely unacceptable for every proposition to be provable and so measures must be taken to prevent this.

---

[i] using fixed point operators exploiting an untyped grammar of mathematical sentences

[ii] Not allowed in Direct Logic because Fix is not allowed.

[iii] Not allowed in Direct Logic because Fix is not allowed..

### Berry's Paradox

Berry's construction [Russell 1906] can be formalized using:

- Least[s] is the smallest integer in the nonempty set of integers s
- Length[s] is length of string s.
- Characterize[s:**StringForPredication**, k:**ℕ**]:**Proposition** ≡
  ∀[x:**ℕ**]→⌊⌈s⌉⌋[x] ⇔ x=k

The above definition of Characterize is *not* allowed in Direct Logic because ⌊⌈s⌉⌋[x] is not a sentence in Direct Logic because ⌈s⌉ is not **Term**◁**Boolean**$^{\sigma}$▷ for some strict type **σ**.

Consider the following definition:

BerryString ≡
"[n:ℕ]→
∀[s:String]→
Length[s]<1000⇨ ¬Characterize[s, n]"[67]

Consider the following set:

BerrySet[68] ≡ {n:**ℕ** | Characterize[BerryString, n]}

Note that if BerrySet existed, then it would not be empty.[69]

1. BerryNumber[70] ≡ Least[BerrySet]
2. Characterize[BerryString, BerryNumber][71]
3. ∀[x:**ℕ**]→⌊⌈BerryString⌉⌋[x] ⇔ x=BerryNumber[72]
4. ⌊⌈BerryString⌉⌋[BerryNumber] ⇔ BerryNumber=BerryNumber[73]
5. ⌊⌈ [n:**ℕ**]→
   ∀[s:**StringForPredication**]→
   Length[s]<1000
   ⇨ ¬Characterize[s, n] ⌉⌋[BerryNumber] ⌋[74]
6. ([n:**ℕ**]→
   ∀[s:**StringForPredication**]→
   Length[s]<100⇨ ¬Characterize[s, n])[BerryNumber]
7. ∀[s:**StringForPredication**]→
   Length[s]<1000 ⇨¬Characterize[s, BerryNumber]
8. Length[BerryString]<1000
   ⇨¬Characterize[BerryString, BerryNumber][75]
9. ¬Characterize[BerryString, BerryNumber][76]

## Sociology of Foundations

> *"Faced with the choice between changing one's mind and proving that there is no need to do so, almost everyone gets busy on the proof."*
> John Kenneth Galbraith [1971 pg. 50]
>
> *Max Planck, surveying his own career in his Scientific Autobiography [Planck 1949], sadly remarked that 'a new scientific truth does not triumph by convincing its opponents and making them see the light, but rather because its opponents eventually die, and a new generation grows up that is familiar with it.'*
> [Kuhn 1962]

Foundations are some ways similar to Complementary Science as defined by [Chang 2007]:

> *[Complementary science] contributes to scientific knowledge through historical and philosophical investigations. [It] asks scientific questions that are excluded from current specialist science. It begins by re-examining the obvious, by asking why we accept the basic truths of science that have become educated common sense. Because many things are protected from questioning and criticism in specialist science, its demonstrated effectiveness is also unavoidably accompanied by a degree of dogmatism and a narrowness of focus that can actually result in a loss of knowledge. History and philosophy of science in its "complementary" mode can ameliorate this situation.*

The inherently social nature of the processes by which principles and propositions in logic are produced, disseminated, and established is illustrated by the following issues with examples:[i]

- **The formal presentation of a demonstration (proof) has not lead automatically to consensus.** Formal presentation in print and at several different professional meetings of the extraordinarily simple proof in this paper have not lead automatically to consensus about the theorem that "Mathematics is Consistent".
- **There has been an absence of universally recognized central logical principles**. Disputes over the validity of the Principle of Excluded Middle led to the development of Intuitionistic Logic, which is an alternative to Classical Logic.

---

[i] *cf.* [Rosental 2008]

- **There are many ways of doing logic.** One view of logic is that it is about *truth*; another view is that it is about *argumentation* (i.e. proofs).
- **Argumentation and propositions have be variously (re-)connected and both have been re-used.** Church's paradox is that assuming theorems of mathematics are computationally enumerable leads to contradiction. In this papers, the paradox is transformed into the fundamental principle that "Mathematics is Open" (*i.e.* it is a theorem of mathematics that the theorems of mathematics are not computationally enumerable) using the argument used in Church's paradox.
- **New technological developments have cast doubts on traditional logical principles.** The pervasive inconsistency of modern large-scale information systems has cast doubt on classical logical principles, *e.g.*, Excluded Middle.[77]
- **Political actions have been taken against views differing from the Establishment Philosophers**. According to [Kline 1990, p. 32], Hippasus was literally thrown overboard by his fellow Pythagoreans "*…for having produced an element in the universe which denied the…doctrine that all phenomena in the universe can be reduced to whole numbers and their ratios*." Fearing that he was dying and the influence that Brouwer might have after his death, Hilbert fired[i] Brouwer as an associate editor of *Mathematische Annalen* because of "*incompatibility of our views on fundamental matters*" [78] *e.g.*, Hilbert ridiculed Brouwer for challenging the validity of the Principle of Excluded Middle.

  Establishment Philosophers have often ridiculed dissenting views and attempted to limit their distribution by political means.[79] Electronic archives and repositories that record precedence of scientific publication in mathematical logic have censored[ii] submissions with proofs such as those in this article.

---

[i] in an unlawful way (Einstein, a member of the editorial board, refused to support Hilbert's action)

[ii] while refusing to provide any justification for the censorship other than administrative fiat

## Appendix 3. Classical Natural Deduction

Below are schemas for nested-box-style Natural Deduction[i] for Classical Mathematics:

| $\Rightarrow$ Introduction | |
|---|---|
| $\Psi$ | ⓘ hypothesis |
| … | |
| $\Phi$ | ⓘ inference |
| $\Psi \Rightarrow \Phi$ | ⓘ **conclusion** |
| $(\Psi \vdash \Phi) \vdash (\Psi \Rightarrow \Phi)$ | |

| $\Rightarrow$ Elimination | |
|---|---|
| $\Psi$ | ⓘ premise |
| … | |
| $\Psi \Rightarrow \Phi$ | ⓘ premise |
| … | |
| $\Phi$ | ⓘ **conclusion** |
| $\Psi, (\Psi \Rightarrow \Phi) \vdash \Phi$ | |

| $\wedge$ Introduction | |
|---|---|
| $\Psi$ | ⓘ premise |
| … | |
| $\Phi$ | ⓘ premise |
| … | |
| $\Psi \wedge \Phi$ | ⓘ **conclusion** |
| $\Psi, \Phi \vdash (\Psi \wedge \Phi)$ | |

| $\wedge$ Elimination | |
|---|---|
| $\Psi \wedge \Phi$ | ⓘ premise |
| … | |
| $\Psi$ | ⓘ **conclusion** |
| … | |
| $\Phi$ | ⓘ **conclusion** |
| $(\Psi \wedge \Phi) \vdash \Psi, \Phi$ | |

| $\vee$ Introduction | |
|---|---|
| $\Psi$ | ⓘ premise |
| … | |
| $\Psi \vee \Phi$ | ⓘ **conclusion** |
| $\Psi \vdash (\Psi \vee \Phi)$ | |

| $\vee$ Elimination | |
|---|---|
| $\neg\Psi$ | ⓘ premise |
| … | |
| $\Psi \vee \Phi$ | ⓘ premise |
| … | |
| $\Phi$ | ⓘ **conclusion** |
| $\neg\Psi, (\Psi \vee \Phi) \vdash \Phi$ | |

| Proof by Contradiction | |
|---|---|
| $\Psi$ | ⓘ hypothesis |
| … | |
| $\Phi \wedge \neg\Phi$ | ⓘ inference |
| $\neg\Psi$ | ⓘ **conclusion** |
| $(\Psi \vdash (\Phi \wedge \neg\Phi)) \vdash \neg\Psi$ | |

| $\vee$ Cases | |
|---|---|
| $\Psi \vee \Phi$ | ⓘ premise |
| … | |
| $\Psi \vdash \Theta$ | ⓘ premise |
| … | |
| $\Phi \vdash \Theta$ | ⓘ premise |
| … | |
| $\Theta$ | ⓘ **conclusion** |
| $(\Psi \vee \Phi), (\Psi \vdash \Theta), (\Phi \vdash \Theta) \vdash \Theta$ | |

| Soundness |
|---|
| $(\vdash \Psi) \Rightarrow \Psi$ |

| Adequacy |
|---|
| $(\Phi \vdash \Psi) \Leftrightarrow (\vdash (\Phi \vdash \Psi))$ |

[i] Evolved from classical natural deduction [Jaśkowski 1934]. See history in Pelletier [1999].

## Index

## End Notes

---

[1] [White 1956, Wilder 1968, Rosental 2008]

[2] Formal typed grammars had not yet been invented when Gödel and other philosophical logicians weakened the foundations of mathematics so that, as expressed, "self-referential" propositions do not infer contradiction. The weakened foundations (based on first-order logic) enabled some nice meta-mathematical theorems to be proved. However, as explained in this article, the weakened foundations are cumbersome, unnatural, and unsuitable as the mathematical foundation for Computer Science.

[3] Mathematical foundations of Computer Science must be general, rigorous, realistic, and as simple as possible. There are a large number of highly technical aspects with complicated interdependencies and trade-offs. Foundations will be used by humans and computer systems. Contradictions in the mathematical foundations of Computer Science cannot be allowed and if found must be repaired.

Classical mathematics is the subject of this article. In a more general context:

- Inconsistency Robust Direct Logic is for pervasively inconsistent theories of practice, e.g., theories for climate modeling and for modeling the human brain.
- Classical Direct Logic can be freely used in theories of Inconsistency Robust Direct Logic. See [Hewitt 2010] for discussion of Inconsistency Robust Direct Logic. Classical Direct Logic for mathematics used in inconsistency robust theories.

[4] Soundness means:

- A theorem of Mathematics can be used *anywhere* including in inconsistency robust inference
- A theorem of Mathematics can be used in a step of a sub-proof to prove a theorem in Mathematics *regardless* of the assumptions of the sub-proof.

[5] Note that this theorem is very different from the result [Kleene 1938], that mathematics can be extended with a proposition asserting its own consistency.

[6] Many of today's most prominent philosophers and logicians have cast doubt on the correctness of the proof.

[7] The definition of inconsistency, *i.e.*,

Consistent$\Leftrightarrow\neg\exists[\Psi$:Proposition$]\rightarrow\ \vdash(\Psi\wedge\neg\Psi)$ is not about numbers.
Consistent with the general practice in Computer Science, there is no way to identify propositions with integers.

[8] A prominent logician referee of this article suggested that if the proof is accepted then consistency should be made an explicit premise of every theorem of classical mathematics!

[9] As shown above, there is a simple proof in Classical Direct Logic that Mathematics ($\vdash$) is consistent. If Classical Direct Logic has a bug, then there might also be a proof that Mathematics is inconsistent. Of course, if a such a bug is found, then it must be repaired.

Fortunately, Classical Direct Logic is simple in the sense that it has just *one* fundamental axiom:

$$\forall[P{:}\mathbf{Boolean}^{\mathbb{N}}]\rightarrow \text{Inductive}[P]\Rightarrow \forall[i{:}\mathbb{N}]\rightarrow P[i]$$
$$\textit{where}\ \forall[P{:}\mathbf{Boolean}^{\mathbb{N}}]\rightarrow$$
$$\text{Inductive}[P] \Leftrightarrow (P[0] \wedge \forall[i{:}\mathbb{N}]\rightarrow P[i] \Rightarrow P[i+1])$$

Of course, Classical Direct Logic has machinery in addition the above axiom that could also have bugs.

The Classical Direct Logic proof that Mathematics ($\vdash$) is consistent is very robust. One explanation is that consistency is built in to the very architecture of classical mathematics because it was designed to be consistent. Consequently, it is not absurd that there is a simple proof of the consistency of Mathematics ($\vdash$) that does not use all of the machinery of Classical Direct Logic.

In reaction to paradoxes, philosophers developed the dogma of the necessity of strict separation of "object theories" (theories about basic mathematical entities such as numbers) and "meta theories" (theories about theories). This linguistic separation can be very awkward in Computer Science. Consequently, Direct Logic does not have the separation in order that some propositions can be more "directly" expressed. For example, Direct Logic can use $\vdash\vdash\Psi$ to express that it is provable that P is provable in Mathematics. It turns out in Classical Direct Logic that $\vdash\vdash\Psi$ holds if and only if $\vdash\Psi$ holds. By using such expressions, Direct Logic contravenes the philosophical dogma that the proposition $\vdash\vdash\Psi$ must be expressed using Gödel numbers.

[10] As shown above, there is a simple proof in Classical Direct Logic that Mathematics ($\vdash$) is consistent. If Classical Direct Logic has a bug, then there might also be a proof that Mathematics is inconsistent. Of course, if a such a bug is found, then it must be repaired.

Fortunately, Classical Direct Logic is simple in the sense that it has one fundamental axiom:

$$\forall[P{:}\mathbf{Boolean}^{\mathbb{N}}]\rightarrow \text{Inductive}[P]\Rightarrow \forall[i{:}\mathbb{N}]\rightarrow P[i]$$
$$\textit{where}\ \forall[P{:}\mathbf{Boolean}^{\mathbb{N}}]\rightarrow \text{Inductive}[P]$$
$$\Leftrightarrow (P[0] \wedge \forall[i{:}\mathbb{N}]\rightarrow P[i] \Rightarrow P[i+1])$$

Of course, Classical Direct Logic has machinery in addition the above axiom that could also have bugs.

The Classical Direct Logic proof that Mathematics ($\vdash$) is consistent is very robust. One explanation is that consistency is built in to the very architecture of classical mathematics because it was designed to be consistent.

In reaction to paradoxes, philosophers developed the dogma of the necessity of strict separation of "object theories" (theories about basic mathematical entities such as numbers) and "meta theories" (theories about theories). This linguistic separation can be very awkward in Computer Science. Consequently, Direct Logic does not have the separation in order that some propositions can be more "directly" expressed. For example, Direct Logic can use $\vdash\vdash\Psi$ to express that it is provable that P is provable in Mathematics. It turns out in Classical Direct Logic that $\vdash\vdash\Psi$ holds if and only if $\vdash\Psi$ holds. By using such expressions, Direct Logic contravenes the philosophical dogma that the proposition $\vdash\vdash\Psi$ must be expressed using Gödel numbers.

11 Classical Direct Logic is different from [Willard 2007], which developed sufficiently weak systems that "self-referential" sentences do not exist.

12 Subsequent further discussion of Wittgenstein's criticism of Gödel's results has unfortunately misunderstood Wittgenstein. For example, [Berto 2009] granted that proof theoretically if $P\Leftrightarrow\nvdash P$, then:

1) $\vdash\nvdash P$

However, the above has proof consequences as follows:

2) $\vdash P$ because $(\nvdash P)\Leftrightarrow P$ in 1) above
3) $\vdash\vdash P$ because of 2) above
4) $\vdash\neg P$ because $(\vdash P)\Leftrightarrow\neg P$ in 3) above

Of course, 2) and 4) are a manifest contradiction in mathematics that has been obtained without any additional "'semantic' story" that [Berto 2009] claimed is required for Wittgenstein's argument that contradiction in mathematics "*is what comes of making up such* ["self-referential"] *sentences*." [Wittgenstein 1956, p. 51e]

13 specified by axioms [Dedekind 1888, Peano 1889] that characterize them up to a unique isomorphism

14 Consequently there is no need to introduce a special kind of set called a "class" that was introduced as a patch for set theory by von Neumann.

15 The *Computational Representation Theorem* [Clinger 1981; Hewitt 2006] characterizes computation for systems which are closed in the sense that they do not receive communications from outside:

> The denotation $\mathsf{Denote}_S$ of a closed system **S** represents all the possible behaviors of **S** as[15]

$$\mathsf{Denote}_S = \lim_{i\to\infty} \mathrm{Progression}_S{}^i$$

$$\textit{where}\ \ \mathrm{Progression}_S{}^i \rightarrow \mathrm{Progression}_S{}^{i+1}$$

In this way, **S** can be mathematically characterized in terms of all its possible behaviors (including those involving unbounded nondeterminism).

The denotations of the Computational Representation Theorem form the basis of procedurally checking programs against all their possible executions.

[16] This is reminiscent of the Platonic divide (but without the moralizing). Gödel thought that "*Classes and concepts may, however, also be conceived as real objects…existing independently of our definitions and constructions*." [Gödel 1944 pg. 456]

[17] Even though English had not yet been invented!

[18] Heuristic: Think of the "elevator bars" $\lfloor ... \rfloor_T$ around s as "raising" the concrete sentence s "up" into the abstract proposition $\lfloor s \rfloor_T$. The elevator bar heuristics are due to Fanya S. Montalvo.

[19] e.g. [Shulman 2012, nLab 2014]

[20] [*cf.* Church 1934, Kleene 1936]

[21] Zermelo in a 1931 letter to Gödel pointed out that in the mathematical theory **ℕ**, there are uncountably many true but unprovable propositions because

- there are uncountably many true propositions in $\{x{=}x \mid x{:}\mathbf{Boolean}^{\mathbb{N}}\}$
- theorems of **ℕ** are countable. Consequently, here is some

  $x_0{:}\mathbf{Boolean}^{\mathbb{N}}$ such that $\models_{\mathbb{N}} x_0 = x_0$ and $\nvdash_{\mathbb{N}} x_0 = x_0$.

[22] Let **Sets◁ℕ⊔ℝ▷** be the closed mathematical theory with axioms for **ℕ**, **ℝ** and **Sets◁ℕ⊔ℝ▷** in this article with **ℕ** a subtype of ℝ. Consequently, $(\vdash_{\mathbf{Sets}\triangleleft\mathbb{N}\sqcup\mathbb{R}\triangleright}\Psi) \Rightarrow\ \vdash\Psi$. Theorems of **Sets◁ℕ⊔ℝ▷** are computational enumerable and it is computationally decidable whether or not a proof is correct in **Sets◁ℕ⊔ℝ▷**

Of course, both of the following hold:

- $\nvdash_{\mathbf{Sets}\triangleleft\mathbb{N}\sqcup\mathbb{R}\triangleright}\mathrm{ChurchTuring}_{\mathbf{Sets}\triangleleft\mathbb{N}\sqcup\mathbb{R}\triangleright}$
- $\nvdash_{\mathbf{Sets}\triangleleft\mathbb{N}\sqcup\mathbb{R}\triangleright}\neg\mathrm{ChurchTuring}_{\mathbf{Sets}\triangleleft\mathbb{N}\sqcup\mathbb{R}\triangleright}$

[23] Consequently, there can cannot be any escape hatch into an unformalized "meta-theory."

[24] The claim also relied on Gödel's "self-referential" proposition.

[25] Formal grammars were invented long after [Gödel 1931].

[26] emphasis in original

[27] **ℕ** is the type of Natural Number axiomatized in this article.

[28] type of 2-element list with first element of type $\boldsymbol{\sigma}_1$ and with second element of type $\boldsymbol{\sigma}_2$

[29] type of term of type $\boldsymbol{\sigma}$

[30] *if* **t** *then* $\boldsymbol{\Phi}_1$ *else* $\boldsymbol{\Phi}_2$

[31] $\boldsymbol{\Phi}_1$, … and $\boldsymbol{\Phi}_k$ infer $\boldsymbol{\Psi}_1$, …, and $\boldsymbol{\Psi}_n$

32 *if* $t_1$ *then* $t_2$ *else* $t_3$

33 Because there is no type restriction, fixed points may be freely used to define recursive procedures on expressions.

34 *if* t *then* $s_1$ *else* $s_1$

35 [Church 1956; Boolos 1975; Corcoran 1973, 1980]. Also, Classical Direct Logic is *not* a univalent homotopy type theory [Awodey, Pelayo, and Warren 2013].

36 $\text{Set}\triangleleft\sigma\triangleright$ is the type of a set of type $\sigma$, $\text{Sets}\triangleleft\sigma\triangleright$ is the type all sets of sets over type $\sigma$, and $\text{Domain}\triangleleft\sigma\triangleright = \sigma\sqcup\text{Sets}\triangleleft\sigma\triangleright$ with the following axioms:

$\{\,\}:\textbf{Set}\triangleleft\sigma\triangleright$ ⓘ the empty set { } is a set

$\forall[x{:}\sigma]\rightarrow \{x\}:\textbf{Set}\triangleleft\sigma\triangleright$ ⓘ a singleton set is a set

$\forall[s{:}\textbf{Sets}\triangleleft\sigma\triangleright]\rightarrow \bigcup s:\textbf{Sets}\triangleleft\sigma\triangleright$ ⓘ all elements of the subsets of a set is a set

$\forall[x{:}\sigma]\rightarrow x\notin\{\,\}$ ⓘ the empty set { } has no elements

$\forall[s{:}\textbf{Set}\triangleleft\sigma\triangleright,\ f{:}\boldsymbol{\sigma}^{\boldsymbol{\sigma}}]\rightarrow (\text{Elementwise}[f])[s]{:}\text{Set}\triangleleft\boldsymbol{\sigma}\triangleright$

ⓘ the function image of a set is a set

$\forall[s{:}\textbf{Set}\triangleleft\sigma\triangleright,\ p{:}\textbf{Boolean}^{\sigma}]\rightarrow s{\restriction}p{:}\textbf{Set}\triangleleft\sigma\triangleright$

ⓘ a predicate restriction of a set is a set

$\forall[s{:}\textbf{Set}\triangleleft\sigma\triangleright]\rightarrow\{\,\}\subseteq s$ ⓘ { } is a subset of every set

$\forall[s_1,s_2{:}\textbf{Set}\triangleleft\sigma\triangleright]\rightarrow s_1=s_2\Leftrightarrow(\forall[x{:}\sigma]\rightarrow x\in s_1\Leftrightarrow x\in s_2)$

$\forall[x,y{:}\sigma]\rightarrow x\in\{y\}\Leftrightarrow x=y$

$\forall[s_1,s_2{:}\textbf{Set}\triangleleft\sigma\triangleright]\rightarrow s_1\subseteq s_2\Leftrightarrow\forall[x{:}\sigma]\rightarrow x\in s_1\Rightarrow x\in s_2$

$\forall[x{:}\sigma;\ s_1,s_2{:}\textbf{Set}\triangleleft\sigma\triangleright]\rightarrow x\in s_1\cup s_2\Leftrightarrow(x\in s_1\vee x\in s_2)$

$\forall[x{:}\sigma;\ s1,s2{:}\textbf{Set}\triangleleft\sigma\triangleright]\rightarrow x\in s1\cap s2\Leftrightarrow(x\in s1\wedge x\in s2)$

$\forall[x{:}\textbf{Domain}\triangleleft\sigma\triangleright;\ s{:}\textbf{Sets}\triangleleft\sigma\triangleright]\rightarrow x\in\bigcup s\Leftrightarrow\exists[s1{:}\textbf{Sets}\triangleleft\sigma\triangleright]\rightarrow x\in s1\wedge s1\in s$

$\forall[y{:}\sigma;\ s{:}\textbf{Set}\triangleleft\sigma\triangleright,\ f{:}\boldsymbol{\sigma}^{\boldsymbol{\sigma}}]\rightarrow y\in(\text{Elementwise}[f])[s]\Leftrightarrow\exists[x\in s]\rightarrow f[x]=y$

$\forall[y{:}\sigma;\ s{:}\textbf{Set}\triangleleft\sigma\triangleright,\ p{:}\textbf{Boolean}^{\sigma}]\rightarrow y\in s{\restriction}p\Leftrightarrow y\in s\wedge p[y]$

The natural numbers are axiomatised as follows where Successor is the successor function:

- $0{:}\mathbb{N}$
- $\text{Successor}{:}\mathbb{N}^{\mathbb{N}}$
- $\forall[i{:}\mathbb{N}]\rightarrow \text{Successor}[i]\neq 0$
- $\forall[i,j{:}\mathbb{N}]\rightarrow \text{Successor}[i]=\text{Successor}[j]\Rightarrow i=j$
- $\forall[P{:}\textbf{Boolean}^{\mathbb{N}}]\rightarrow \text{Inductive}[P]\Rightarrow\forall[i{:}\mathbb{N}]\rightarrow P[i]$

  where $\forall[P{:}\textbf{Boolean}^{\mathbb{N}}]\rightarrow$

  $\text{Inductive}[P]\Leftrightarrow P[0]\wedge\forall[i{:}\mathbb{N}]\rightarrow P[i]\Rightarrow P[\text{Successor}[i]]$

37 *I.e.*, $\nexists[s{:}\textbf{Sets}\triangleleft\mathbb{N}\triangleright]\rightarrow\forall[e{:}\textbf{Domain}\triangleleft\mathbb{N}\triangleright]\rightarrow e\in s\Leftrightarrow e{:}\textbf{Sets}\triangleleft\mathbb{N}\triangleright$ where $\textbf{Domain}\triangleleft\mathbb{N}\triangleright=\mathbb{N}\sqcup\textbf{Sets}\triangleleft\mathbb{N}\triangleright$

38 a set is not well founded if and only if it has an infinite ∈ chain

39 [Dedekind 1888, Peano 1889]

[40] [Dedekind 1888, Peano 1889]

[41] ℕ is identified with the type of natural numbers

[42] which can be equivalently expressed as:

∀[P:**Boolean**$^{ℕ}$]→ Inductive[P]⇨ ∀[i:**ℕ**]→ P[i]=True

*where*

∀[P:**Boolean**$^{ℕ}$]→

Inductive[P] ⇔ (P[0]=True ∧ ∀[i:]→ P[i]=True ⇨P[i+1]=True)

[43] [Dedekind 1888]

[44] ℝ is identified with the type of natural numbers

[45] *cf.* [Zermelo 1930].

[46] The Continuum Hypothesis remains an open problem for Direct Logic because its set theory is very powerful. The forcing technique used to prove the independence of the Continuum Hypothesis for first-order set theory [Cohen 1963] does not apply to Direct Logic because of the strong induction axiom [Dedekind 1888, Peano 1889] used in formalizing the natural numbers **ℕ.**

Of course, trivially,

(⊨$_{Domain◁ℕ▷}$ContinuumHypothesis)∨(⊨$_{Domain◁ℕ▷}$¬ContinuumHypothesis)

where Domain◁σ▷≡σ⊔**Sets**◁σ▷.

[47] [Wittgenstein in 1937 published in Wittgenstein 1956, p. 50e and p. 51e]

[48] [Wang 1997] pg. 197.

[49] In 1666, England's House of Commons introduced a bill against atheism and blasphemy, singling out Hobbes' Leviathan. Oxford university condemned and burnt Leviathan four years after the death of Hobbes in 1679.

[50] There is no subset of ℝ whose cardinality is strictly between ℕ and ℝ.

[51] [Cohen 2006] Cohen's proof was a great achievement in spite of the weakness of his theorem.

[52] because the computable real numbers are enumerable.

[53] [Zermelo 1930, van Dalen 1998, Ebbinghaus 2007]

[54] First-order theories fall prey to paradoxes like the Löwenheim–Skolem theorems (*e.g.* any first-order theory of the real numbers has a countable model). First-order theorists have used the weakness of first-order logic to prove results that do not hold in stronger formalisms such as Direct Logic [Cohen 1963, Barwise 1985].

[55] *e.g.* the theory *RealClosedField*[Tarski 1951]

[56] i.e. in classical notation: {¬P[m] | m:**ℕ**}

[57] unique up to isomorphism via a unique isomorphism

[58] Unlike Direct Logic, which is more expressive because propositions are not countable.

[59] As pointed out elsewhere in this paper, in the more powerful system of Direct Logic, Löb's theorem when generalized to all of mathematics turns

into a paradox because Direct Logic has the Principle of Integrity that in mathematics: (⊢Φ)⇨Φ, which does not result in the same proof of contradiction because "self-referential" propositions are not allowed in Direct Logic.

[60] If the principle were allowed, then 𝒫ℒ would be inconsistent because every sentence would be provable in 𝒫ℒ by Löb's theorem.

[61] [Tarski and Vaught 1957]

[62] Statement of Church's Paradox

[63] Unfortunately, in formalizing Gödel's proof, [Shankar 1994] and [O'Connor 2005] followed Gödel in using integers to code "self-referential" sentences using fixed points (exploiting an untyped grammar of sentences). According to [Cantini 2012]:

> In the twenties and in the early thirties, the orthodox view of logic among mathematical logicians was more or less type- or set theoretic. However, there was an effort to develop new grand logics as substitutes for the logic of Principia Mathematica. These frameworks arose both as attempts to recover the simplicity of the type-free approach, as derived from the so-called naive comprehension principle, as well as in order to satisfy meta-mathematical needs, such as the clarification of fundamental concepts underlying the notions of "formal system," "formalism," "rule," *etc....*
>
> However, the theories of Curry and Church were almost immediately shown inconsistent in 1934, by Kleene and Rosser, who (essentially) proved a version of the Richard paradox (both systems can provably enumerate their own provably total definable number theoretic functions). The result was triggered by Church himself in 1934, when he used the Richard paradox to prove a kind of incompleteness theorem (with respect to statements asserting the totality of number theoretic functions)....
>
> The reason for the inconsistencies was eventually clarified by Curry's 1941 essay. There he distinguishes two basic completeness notions: a system *S* is deductively complete, if, whenever it derives a proposition B from the hypothesis A, then it also derives the implication A⇨B (deduction theorem or introduction rule for implication); *S* is combinatorially complete[63] if, whenever M is a term of the system possibly containing an indeterminate x, there exists a term (Church's λ[x]→ M) naming the function of x defined by M. Curry then remarks that the paradox of Kleene-Rosser arises because Church's and Curry's systems satisfy both kinds of completeness, thus showing that the *two properties are incompatible*.

[64] See Historical Appendix.

[65] In Direct Logic, fixed points on propositions cannot be shown to be propositions and consequently the proofs are not valid for the following reason:

The fixed point operator cannot be defined using the typed lambda calculus:

PropToProp ≡ [Proposition]↦Proposition

Helper▪[f:PropToProp]:PropToProp ≡
[x:([?]↦PropToProp)]→ f[x▪[x]]

Fix▪[f:([PropToProp]↦PropToProp)]:PropToProp ≡
(Helper▪[f])▪[Helper▪[f]]

The missing strict type ? does not exist.

[66] Recently, [Yanofsky 2013 page 328] has expressed concern about Löb's paradox:

> *we must restrict the fixed-point machine in order to avoid proving false statements* [using Löb's argument]. *Such a restriction might seem strange because the proof that the fixed-point machine works seems applicable to all* [functions on sentences in an untyped grammar of sentences]. *But restrict we must.*

Yanofsky solved the above problem posed by Löb's paradox using systems of logic that are so weak that they cannot abstract their own sentences. Unfortunately, such weak systems are inadequate for Computer Science. Instead of weakening, Direct Logic adopted the strategy of barring "self-referential" propositions by using a grammar for sentences that does not allow the fixed-point machinery for sentences.

[67] Note that Length[BerryString]<100

[68] Note that BerrySet is not allowed in Direct Logic because Characterize is not allowed.

[69] Consider the following definition using ε notation of Hilbert:

TrivialCharacterization[k:ℕ] ≡
String["[x:ℕ]→ (ε[y:ℕ]→ y=",
IntegerToString[k],
") =",
IntegerToString[k]]

Clearly, ∀[k:ℕ]→ Characterize[TrivialCharacterization[k], k].
Consequently, Characterize[TrivialCharacterization[0], 0]
Note that length[TrivialCharacterization[0]]<1000. Thus BerrySet is not empty.

[70] Note that BerryNumber is not defined in Direct Logic because BerrySet is not defined.

[71] using definition of BerryNumber

[72] using definition of Characterize

[73] substituting BerryNumber for x

[74] using definition of BerryString

[75] substitution of BerryString for s

[76] Contradicting step 2.

[77] for discussion see [Hewitt 2010]

[78] Hilbert letter to Brouwer, October 1928

[79] *e.g.* "*The problem with such papers* [critiquing Establishment doctrine] *is that casual readers will use them to criticize and maybe stop future funding ...*" [Berenji, *et. al.*, 1994]